\documentclass[twocolumn,showpacs,amsmath,amssymb,pra,nofootinbib]{revtex4}

\usepackage{graphicx}
\usepackage{dcolumn}
\usepackage{bm}
\usepackage{times,mathptm}

\newcommand{\ket}[1]{|#1\rangle}
\newcommand{\bra}[1]{\langle#1|}

    {
    \smallskip
    \refstepcounter{theorem}
    \noindent
    {\bf Example \Alph{section}.\arabic{theorem}} \ \ }
    {\hspace*{\fill}{$\Diamond$}
    \smallskip}

\newenvironment{appremark}
    {
    \smallskip
    \refstepcounter{theorem}
    \noindent
    {\bf Remark \Alph{section}.\arabic{theorem}} \ \ }
    {\hspace*{\fill}{$\Diamond$}
    \smallskip}

\newenvironment{appdefinition}
    {
    \smallskip
    \refstepcounter{theorem}
    \noindent
    {\bf Definition \Alph{section}.\arabic{theorem}} \ \ }
    {\hspace*{\fill}{\ }
    \smallskip}

    {
    \smallskip
    \refstepcounter{theorem}
    \noindent
    {\bf Example \arabic{section}.\arabic{theorem}} \ \ }
    {\hspace*{\fill}{$\Diamond$}
    \smallskip}

    {
    \smallskip
    \refstepcounter{theorem}
    \noindent
    {\bf Remark \arabic{section}.\arabic{theorem}} \ \ }
    {\hspace*{\fill}{$\Diamond$}
    \smallskip}

    {
    \smallskip
    \refstepcounter{theorem}
    \noindent
    {\bf Definition \arabic{section}.\arabic{theorem}} \ \ }
    {\hspace*{\fill}{\ }
    \smallskip}

    {
    \smallskip
    \refstepcounter{theorem}
    \noindent
    {\bf Definition \Alph{section}.\arabic{theorem}} \ \ }
    {\hspace*{\fill}{\ }
    \smallskip}

    {
    \smallskip
    \refstepcounter{theorem}
    \noindent
    {\bf Scholium \arabic{section}.\arabic{theorem}} \it \ \ }
    {\hspace*{\fill}{\ }
    \smallskip}

\newenvironment{proof}[1][]
    {
    \noindent
    {\bf Proof{#1}:  }
    }
    {\hspace*{\fill}{$\Box$}\smallskip}

\newenvironment{sketch}[1][]
    {
    \noindent
    {\bf Sketch{#1}:  }
    }
    {\hspace*{\fill}{$\Box$}\smallskip}

    {
    \noindent
    {\bf Reference:  }
    }
    {\hspace*{\fill}{$\odot$}\smallskip}

\newtheorem{theorem}{Theorem}[section]
\newtheorem{proposition}[theorem]{Proposition}
\newtheorem{lemma}[theorem]{Lemma}

\begin{document}
\bibliographystyle{apsrev}

\title{Stability of global entanglement in thermal states of spin chains}

\author{Gavin K. Brennen$^{1}$}\email{gavin.brennen@nist.gov}
\author{Stephen S. Bullock$^2$} \email{stephen.bullock@nist.gov}

\affiliation{
$^1$ National Institute of Standards and Technology, Atomic Physics Division,
Gaithersburg, Maryland, 20899-8420 \\
$^2$ National Institute of Standards and Technology, Mathematical and 
Computational Sciences Division, Gaithersburg, Maryland 20899-8910 \\
}

\setlength{\unitlength}{0.125in} 

\begin{abstract}
We investigate the entanglement properties of a one dimensional chain of qubits coupled via nearest neighbor spin-spin interactions.  The entanglement measure used is the $n$-concurrence, which is distinct from other measures on spin chains such as bipartite entanglement in that it can quantify ``global" entanglement across the spin chain.  Specifically, it computes the overlap of a quantum state with its time-reversed state.   As such this measure is well suited to study ground states of spin chain Hamiltonians that are intrinsically time reversal symmetric.  We study the robustness of $n$-concurrence of ground states when the interaction is subject to a time reversal antisymmetric magnetic field perturbation.  The $n$-concurrence in the ground state of the isotropic XX model is computed and it is shown that there is a critical magnetic field strength at which the entanglement experiences a jump discontinuity from the maximum value to zero.  The $n$-concurrence for thermal mixed states is derived and a threshold temperature is computed below which the system has non zero entanglement. 
\end{abstract}
\maketitle
\section{Introduction}
There is considerable interest in understanding the distinction between quantum and classical correlations in many-body systems.  Common examples found in nature are collections of spins coupled by pairwise interactions.   Spin chain Hamiltonians are described by nearest neighbor interactions between spins (usually $s=1/2$) particles.  A typical example is the one dimensional quantum XYZ model: 
\begin{equation}
H_{XYZ}(h)=\sum_j J_x\sigma^x_j\sigma^x_{j+1}+J_y\sigma^y_j\sigma^y_{j+1}+J_z\sigma^z_j\sigma^z_{j+1}+h\sigma^z_j.\;
\end{equation}
which describes pairwise interactions with a homogeneous external magnetic field $h$ that acts locally on each spin. 
While these models are only an approximation to the real physics, they are rich enough to extract essential statistical properties of the underlying system.  A striking phenomenon in many of these systems is the existence of a quantum phase transition (QPT) described by the analytic discontinuity in some thermodynamic quantity with the variation of a system interaction parameter.  For the XY model ($J_z=0$) the QFT is manifested by the divergence in the range of pairwise spin correlations at a critical magnetic field strength.

Given that long range classical correlations are present in the ground states of spin chains it is to be expected that these correlations can be related to functions quantifying quantum entanglement.  Several significant results have already been obtained for the XY model.  It has been shown that a measure of entanglement between pairs of spins, the $2$-concurrence, experiences a QFT at the same critical magnetic field strength as the transition point for classical spin correlations \cite{Nielsen,Osterloh}.  These studies also show that in the isotropic, or XX model ($J_x=J_y,J_z=0$), the range of pairwise entanglement is infinite in the thermodynamic limit even though this case does not admit a QFT.   The entanglement in a bipartite division of a spin chain into two contiguous blocks has been studied in Ref. \cite{GVidal:03}.  There the von Neumann entropy of the reduced state of one block was used as the entanglement measure.  By this measure, that the entanglement obeys universal scaling laws in accordance with conformal invariance.  In another work, it was proven that there is another characterization of entanglement, the localizable entanglement $\xi_E$, whose range is always at least as long as classical correlation lengths \cite{Verstraete:04}.  The quantity $\xi_E$ is defined as the maximum pairwise entanglement that can be localized on two qubits, on average, by optimizing over local operations on the other qubits.  

These investigations focused on bipartite entanglement between individual spins or blocks of spins.  Recently, it was shown that the notion of entanglement can be generalized beyond subsystems by computing the purity of state with respect to a chosen subalgebra \cite{Barnum:03}.  Applied to the XY model, the relevant subalgebra is the set of number non-conserving fermionic operators that connect different irreducible representations of the unitary Lie algebra $\mathfrak{u}(n)$, where $n$ is the number of modes.  The purity can then be expressed in terms of fluctuations of the total fermion number and it is characterized by a second order QFT at the critical magnetic field strength \cite{Somma:04}.

All of the aforementioned results add considerable insight into how nonlocal correlations between spins are distributed in the ground states of spin chain Hamiltonians.  In this paper we take a different approach to studying the entanglement of spin chains.  We do not seek a correspondence in the behavior of entanglement and classical correlation functions near critical points.  Rather, we investigate the stability of  ``global" entanglement across the system when subject to environmentally influenced effects such as finite temperature and perturbation by a magnetic field.  Any approach to this problem must contend with the non-uniqueness of a single measure of global entanglement in a multipartite system.  We pick a measure of entanglement, the $n$-concurrence, that is motivated by an underlying time reversal symmetry of the XYZ model.  The hope is that by studying the behavior of global entanglement we can gain some insight into how long range quantum correlations appear in systems that are already highly correlated classically.  

The paper is organized as follows.  The properties of the entanglement measure are discussed in Sec. \ref{sec:prop}.  In Sec. \ref{sec:gse} the ground state entanglement in the one dimensional XY model with periodic boundaries is studied.  It is shown that the XX model displays a jump discontinuity in $n$-concurrence when the perturbing magnetic field reaches a critical strength.  The entanglement in a thermal state is computed in Sec. \ref{sec:therm} and the threshold temperature is derived which sets an upper bound on how mixed the state can be before the $n$-concurrence vanishes.  The computation of entanglement for thermal states relies on an important theorem derived in Appendix \ref{app} that yields a closed form expression for the concurrence of mixed states.  In Sec. \ref{sec:openbndr} the preceding analysis is extended to the quantum XX model with open boundaries.  Issues concerning the experimental observation of entanglement  in spin chains are discussed in Sec. \ref{sec:meas}.  Finally, a summary and conclusions are presented in Sec. \ref{sec:concl}.

\section{Properties of $n$-concurrence}
\label{sec:prop}
A basic mathematical tool for studying entanglement is the entanglement monotone.  It is a mathematical function that maps states to real numbers and exhibits two important properties.  First, it is zero for separable states, i.e. those that can be described purely by classical probability distributions; second, it is non-increasing on average under local operations and classical communication.  There are many monotones to choose from that quantify entanglement in multipartite systems.  The choice of measure may be best dictated by the underlying symmetries of the system if there are any.

The monotone we study is concurrence.  The $2$-concurrence was originally derived by Wootters \cite{Wootters:98} and later generalized to any even number of qubits $n$ \cite{Wong:01}. On a pure state $\ket{\psi}$ the $n$-concurrence is defined
\begin{equation}
C_n( \ket{\psi} )\ =\ \big| \bra{\psi}\mho\ket{\psi} \big|,
\end{equation}
where $\mho$ is an anti-unitary time-reversal operator.  When acting on $n$ qubits, we can write $\mho=[\prod_{j=1}^n(-i\sigma_j^y)]\tau$, where $\tau$ is the complex conjugation operator.  The $n$-concurrence and its square, the $n$-tangle, have been shown to be entanglement monotones \cite{BB:03, Wong:01} for $n$ even, and for $n$ odd are identically zero.  The range of the measure is $0\leq C_n(\ket{\psi})\leq 1$.  

The $n$-concurrence is an attractive measure for two reasons.  First, it is sensitive to global entanglement in the sense that it is zero if any qubit is separable from the rest of the system.  This does not imply that each qubit is entangled with every other qubit, however.  For instance, the $n$-party Greenberger-Horne-Zeilinger (GHZ) state $\ket{\mbox{GHZ}_n}=1/\sqrt{2}(\ket{0\ldots 0}+\ket{1\ldots 1})$ is maximally concurrent but this can also be the case for sub-global entanglement, e.g. $C_8(\ket{\mbox{GHZ}_{4}}\otimes\ket{\mbox{GHZ}_{4}})=1$.  Furthermore, some entangled states have vanishing concurrence, e.g. $\ket{W_4}=(1/2)[\; \ket{0001}+\ket{0010}+\ket{0100}+\ket{1000} \; ]$.  Nevertheless, it is sensitive to entanglement described by superpositions of states and their spin flips.  Second, the $n$-concurrence measures the overlap of a quantum state with its time reversed state and it is thus a natural function to choose for eigenstates of a Hamiltonian that respects this symmetry.  Any Hamiltonian that can be written as a sum of tensors of an even number of non identity Pauli operators only is time reversal symmetric.  For example, exclusively pairwise interactions satisfy this condition
\begin{equation}
\mho H_{XYZ}(0)\mho^{-1}\ = \ H_{XYZ}(0).
\end{equation}
This symmetry has important consequences with regard to entanglement of eigenstates.
\begin{proposition}[\cite{BBOL:04}]
\label{prop:us}
Let $H$ be a Hamiltonian on some number $n$ of quantum-bits.
Suppose $H$ has time-reversal symmetry with respect to $\mathbb{f}$.
Let $\lambda$ be a fixed eigenvalue of $H$.
Then either (i) $\lambda$ is degenerate
(i.e. has multiplicity at least two,) or 
(ii) the normalized eigenstate $\ket{\lambda}$
has $C_n(\ket{\lambda})=1$.  Should $n=2p-1,\ p\in{\mathbb N}$, then case (i) always holds.
\end{proposition}

Notice that if any spin does not interact with the others in a collection of spins then there will be a degeneracy.  There are several examples of spin-chain Hamiltonians with non-degenerate ground states.  Among them is the XYZ Hamiltonian with $J_x=J_y=J_z\equiv J>0$, also known as the Heisenberg interaction.   It's ground state was proven by Lieb and Mattis \cite{Lieb:62} to be non-degenerate in any number of dimensions, with or without periodic boundary conditions provided the underlying lattice has reflection symmetry about some plane.  Additionally, in 1D, the XY Hamiltonians with $J_x,J_y\neq 0,J_z=0$ are non-degenerate \cite{Katsura}.  In this paper we focus on entanglement properties of the latter.

Extending measures of entanglement on pure states to ensembles of pure states, or mixed states, can be carried out by averaging over the entanglement of pure states in the ensemble.  The choice of the state decomposition should not increase the amount of entanglement, therefore, the entanglement should be minimized over all valid decompositions of the state.  This formulation is known as the convex roof of the function and for $n$-concurrence it is written:
\begin{equation}
C_n(\rho)=\min \bigg\{\sum_j p_jC_n(\ket{\psi^j});\ \rho=\sum_j p_j\ket{\psi^j}\bra{\psi^j}\bigg\}.\;
\end{equation}
Remarkably, the $n$-concurrence of a mixed state on $n$ even qubits can be expressed in closed form (see Appendix \ref{app}):
\begin{equation}
C(\rho)=\max \bigg\{0,\lambda_0-\sum_{j=1}^{N-1}\lambda_j\bigg\}.\;
\label{concmixed}
\end{equation}
Here $\{\lambda_j\}_{j=0}^{N-1} \; = \; \mbox{spec}(\sqrt{\rho \mho\rho\mathbb{f}^{-1}})$, where $\mbox{spec}(A)$ denotes
the spectrum of the operator $A$, and $N=2^n$ is the dimension of the system.  The set of eigenvalues are real, positive numbers arranged in non-increasing order: $\lambda_0\geq \lambda_1 \ldots\geq \lambda_{N-1}$.

\section{Ground state entanglement}
\label{sec:gse}

The XY model with a uniform magnetic field $h$ is written
\begin{equation}
H_{XY}(h)=\sum_{l=1}^{n} J_x\sigma_l^x\sigma^x_{l+1}+J_y \sigma^y_l\sigma^y_{l+1}+h \sigma_1^z.\;
\label{HXYfirst}
\end{equation}
Here we have assumed cyclic boundary conditions $(\sigma^{\alpha}_{n+k}\equiv \sigma^{\alpha}_k)$.  Additionally, we assume $n$ to be even.  The presence of the magnetic field breaks the time reversal symmetry of the interaction.  Indeed, the magnetic field interaction $H_B=h\sum_{j=1}^n\sigma_j^z$ contains a sum of terms involving only odd numbers of Pauli operators and therefore it is time-reversal anti-symmetric \cite{BBOL:04}: $\mho H_B\mho^{-1}=-H_B$, i.e. 
\begin{equation}
\mho H_{XY}(h)\mho^{-1}=\; H_{XY}(-h).\;
\label{reversed}
\end{equation}
We study how robust the $n$-concurrence of the ground state is to perturbation by the magnetic field.

The eigenvalues and eigenvectors of $H_{XY}(h)$ can be solved for exactly by performing the Jordan-Wigner transformation from Pauli operators to fermionic operators followed by a Fourier transformation.  The Jordan-Wigner transformation is given by 
\begin{equation}
a_j^{\dagger}=\nu_j \sigma^-_j,\quad a_j=\sigma^+_j\nu_j ,\;
\end{equation}
where the introduction of the non-local terms $\nu_l=\otimes_{k=1}^{l-1}\sigma^z_k=\otimes_{k=1}^{l-1}(1-2n_k)=(-1)^{\sum_{k=1}^{l-1}a_k^{\dagger}a_k}$ enforces the anticommutation relations:
\begin{equation}
\{a_k^{\dagger},a_{k^{\prime}}\}=\delta_{kk^{\prime}},\quad \{a_k,a_{k^{\prime}}\}=\{a_k^{\dagger},a_{k^{\prime}}^{\dagger}\}=0.\;
\end{equation}
Define the Fourier transformed creation and annihilation operators as:
\begin{equation}
A_j^{\dagger}\ =\ \frac{1}{\sqrt{n}}\sum_{k=1}^na_k^{\dagger}e^{i\pi (jk/n-1/4)}
\end{equation}
where $A_{2 n+k}^{\dagger}\equiv A_k^{\dagger}$.
Following Katsura \cite{Katsura} we can decompose $H_{XY}(h)$ into two subspaces corresponding to the $\pm$ eigenvalues of the operator $\nu_{n+1}=\otimes_{k=1}^{n}\sigma_k^z=
(-1)^{\sum_{k=1}^{n}a_k^{\dagger}a_k}=(-1)^{\sum_{k=1}^{n}A_{2k}^{\dagger}A_{2k}}$.  This amounts to a partitioning of the system into even and odd parity halves of the total number of excitations: 
\begin{equation}
H_{XY}(h)=\frac{1}{2}(1+\nu_{n+1})H^+ +\frac{1}{2}(1-\nu_{n+1})H^-.\;
\end{equation}
In terms of the new fermionic operators $A_k,A_k^{\dagger}$, the Hamiltonian $H_{XY}$ reduces to block diagonal form with each block spanned by particle occupation in positive and negative ``momentum'' number states $A_k^{\dagger}A_k$.  The even parity sector of $H_{XY}$ is
\begin{equation}
H^+\ =\ \sum_{k=1}^{n/2}H_{2k-1}\;
\end{equation}
where 
\begin{equation}
H_{k}\; = \; 
\left(\begin{array}{rrrr}
-2\epsilon_k & -2\delta_k & 0 & 0 \\
-2\delta_k & 2\epsilon_k & 0 & 0\\
0 & 0 & 0 & 0 \\
0 & 0 & 0 & 0 \\
\end{array}
\right)\qquad \mbox{for}\ 1\leq k\leq n,
\end{equation}
in the basis $\{\ket{0},A_k^{\dagger}A_{-k}^{\dagger}\ket{0},A_k^{\dagger}\ket{0},A_{-k}^{\dagger}\ket{0}\}$.  The matrix elements are 
\begin{equation}
\epsilon_k=(J_x+J_y)\cos(\pi k/n)-h,\quad \delta_k=(J_x-J_y)\sin(\pi k/n).
\end{equation}
Similarly, the odd parity sector is, 
\begin{equation}
H^-\; =\; H_0+H_n+\sum_{k=1}^{n/2-1}H_{2k}
\end{equation}
where the two boundary terms are
\begin{equation}
H_0\; =\; \left(\begin{array}{rr}
-\epsilon_0 & 0\\
0 & \epsilon_0 \\
\end{array}
\right),\quad H_n\; =\; \left(\begin{array}{rr}
-\epsilon_n& 0\\
0 & \epsilon_n\\
\end{array}
\right)
\end{equation}
in the bases $\{\ket{0},A_0^{\dagger}\ket{0}\}$ and $\{\ket{0},A_n^{\dagger}\ket{0}\}$ respectively.

The Hamiltonian can be brought to full diagonal form by a Bogoliubov transformation on the operators $H_k$ to a set of fermionic operators $\beta_k,\beta_k^{\dagger}$ whose number is conserved.  The transformation is:
\begin{equation}
\beta_k=\cos (\theta_k) A_k+\sin (\theta_k) A_{-k}^{\dagger},\quad \beta_k^{\dagger}=\cos (\theta_k) A_k^{\dagger}+\sin (\theta_k) A_{-k}
\end{equation}
where $\tan(2\theta_k)=\delta_k/\epsilon_k$.
In many treatments of the statistical properties of the XY model, the boundary terms $H_0,H_n$ are dropped with the justification that their contribution is negligible in the thermodynamic limit \cite{Lieb:61}.  Because we are interested in entanglement properties of the system for all even $n$, we keep these terms. 

\begin{figure}
\begin{center}
\includegraphics[scale=0.21]{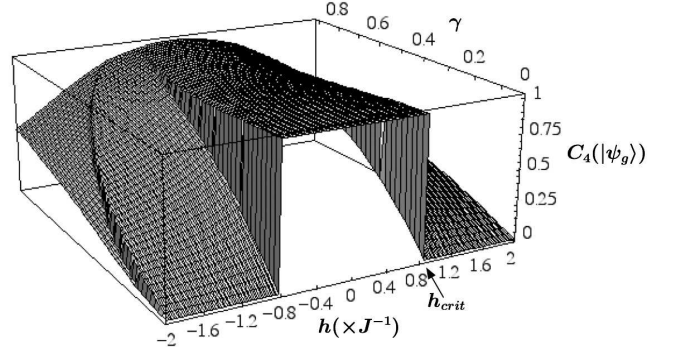}
\caption{\label{fig:1}Plot of the $4$-concurrence of the XY model as a function of magnetic field strength and anisotropy $\gamma=(J_x-J_y)/2J$ where $J=(J_x+J_y)/2$.  In the isotropic case $(J_x=J_y)$, there is a critical magnetic field strength $h_{crit}$ where the $n$-concurrence experiences a jump discontinuity from a value of one for $|h|<h_{crit}$ to zero at $|h|\geq h_{crit}$.  For $n=4$, $h_{crit}=2J\tan(\pi/8)$.}
\end{center}
\end{figure}

We wish to compute the $n$-concurrence of the ground state of $H_{XY}(h)$ which we denote $\ket{\psi_g(h)}$. 
One approach is to explicitly calculate $C_n(\ket{\psi_g(h)})=|\bra{\psi_g(h)}\mho\ket{\psi_g(h)}|$.  Because the Hamiltonian is real, the energy eigenstates are real and this amounts to computing the expectation value $\bra{\psi_g}(-i\sigma^y)^{\otimes n}\ket{\psi_g}$.  After performing the transformation from the operators $\sigma_j^y$ to the fermionic operators, there will be $O(n^2)$ terms to sum in the expectation value.  An alternative approach is facilitated using the expression for the $n$-concurrence of mixed states, Eq. \ref{concmixed}.  The problem is reduced to finding $\mbox{spec}(\sqrt{\rho\mho\rho\mho^{-1}})$, associated with the ground state:
\begin{equation}
\rho_{T=0}\; =\; \ket{\psi_g(h)}\bra{\psi_g(h)}\; =\; \lim_{T\to 0}\frac{e^{-\beta H_{XY}(h)}}{Z},.
\end{equation}
where $Z=\mbox{Tr}[e^{-\beta H}]$ is the partition function.  In describing the state as the $T\rightarrow 0$ limit of the thermal sample we have implicitly assumed the ground state is non-degenerate.  All ground states of $H_{XY}(0)$ with $J_x,J_y\neq 0$ are non degenerate but degeneracies do occur at finite magnetic field strengths.  In the present discussion we assume that at any given degeneracy point, the state of the system is in a single (zero entropy) pure state.  This assumption is dropped in Sec. \ref{sec:therm} where it is shown that thermal states with degenerate ground states have zero $n$-concurrence.  Proceeding with this qualification in mind we have
\begin{equation}
\begin{array}{lll}
\mho\rho_{T=0}\mho^{-1}&=&\mho\bigg({\displaystyle \lim_{T\to 0}}\frac{e^{-\beta H_{XY}(h)}}{Z}\bigg)\mho^{-1}\\
&=&{\displaystyle \lim_{T\to 0}}\mho\frac{e^{-\beta  H_{XY}(h)}}{Z}\mho^{-1}\\
&=&{\displaystyle \lim_{T\to 0}}\frac{e^{-\beta \mho H_{XY}(h)\mho^{-1}}}{Z}\\
&=&{\displaystyle \lim_{T\to 0}}\frac{e^{-\beta H_{XY}(-h)}}{Z}\\
&=&\ket{\psi_g(-h)}\bra{\psi_g(-h)},\;
\end{array}
\end{equation}
where in the fourth line we have used Eq. \ref{reversed}.
The matrix $\rho_{T=0}\mho\rho_{T=0}\mho^{-1}$ is at most rank one and the $n$-concurrence of the ground state of $H_{XY}(h)$ is therefore,
\begin{equation}
C_n(\ket{\psi_g(h)})\ =\ \big| \bra{\psi_g(h)}\psi_g(-h)\rangle \big|.
\label{concHXY}
\end{equation}

The entanglement of the ground state of the XY model in $n=4$ qubits is plotted in Fig. \ref{fig:1}.  Notice the symmetry of the $n$-concurrence under $h\rightarrow -h$ as anticipated in Eq. \ref{concHXY}.  At zero magnetic field $H_{XY}$ is time reversal symmetric and non-degenerate so the $4$-concurrence is equal to one for anisotropies in the range $0\leq \gamma < 1$.  In the isotropic case, there appears to be a jump discontinuity in the entanglement at a critical magnetic field strength $h_{crit}$.  We proceed to show that this phenomenon arises for all even $n$ and that $h_{crit}$ is given by the minimum field strength at which the ground eigenstate becomes doubly degenerate.  
 
Henceforth, we focus on the isotropic XX model $(J_x=J_y\equiv J)$.  In this case, the Hamiltonian $H_{XX}(h)$ is already diagonal in the number operators $A_k^{\dagger}A_k,A_{-k}^{\dagger}A_{-k}$.    
The eigenvalues of $H^+$ are given by:
\begin{equation}
E^+=\bigg\{\sum_{k=1}^{n/2}x_{2k-1};\quad x_{k}\in\{-\epsilon_{k},\epsilon_{k},0,0\}\bigg\}\;
\end{equation}
There are a total of $4^{n/2}$ eigenvalues obtained by the sum over elements in the set.  However, the projection $1/2(1+\nu_{n+1})H^+$ requires that each energy eigenvalue be equal to a sum of $n/2$ terms in the set with the difference between the number of $\pm$ signs in the sum satisfying $\#(+)-\#(-)=n/2\mod 2$.   This makes the total number of eigenvalues obtained from $H^+$ equal to $2^n/2$. 
Similarly,
\begin{equation}
E^-=\bigg\{\pm\epsilon_0\pm\epsilon_n+\sum_{k=1}^{n/2-1}x_{2k};\quad x_{k}\in\{-\epsilon_{k},\epsilon_{k},0,0\}\bigg\},\;
\end{equation}
where the projection $1/2(1-\nu_{n+1})H^-$ requires that each energy eigenvalue be equal to a sum of $n/2$ terms satisfying $\#(+)-\#(-)=(n/2+1)\mod 2$. 

To analyze the concurrence of the ground states we note that $H_{XX}(h)$ and the $z$ projection of the collective spin operator, $S_z=\sum_{j=1}^n\sigma^z_j$, are simultaneously diagonalizable, i.e.
\begin{equation}
\bigg[\ \sum_{j=1}^nJ\sigma^x_j\sigma^x_{j+1}+J\sigma^y_j\sigma^y_{j+1}+hS_z
\; ,\; S_z\ \bigg]=0.\;
\end{equation} 
This is not true for the anisotropic interaction.  The fact that $S_z$ is a conserved quantity implies that while the eigenvalues of $H_{XX}(h)$ are functions of $h$, the eigenstates are independent of $h$.  We label the (possibly degenerate) ground state of $H_{XX}(h)$ with its corresponding eigenvalues $s_z$ of $S_z$, as $\ket{\psi_g^{s_z}(h)}$.  Notice the expectation value of $S_z$ in terms of the fermionic number operators is
\begin{equation}
\begin{array}{lcl}
\big\langle S_z \big\rangle & \; = \; & \big\langle\;
\sum_{k=1}^n(1-2a_k^{\dagger}a_k)\; \big\rangle\\
&\; =\; &n-2 \big\langle\;\sum_{k=1}^nA_{2k}^{\dagger}A_{2k}\; \big\rangle\\
&\; =\; &n-2\big\langle\;\sum_{k=1}^nA_{2k}^{\dagger}A_{2k}\; \big\rangle\;
\end{array}
\end{equation}
Introducing the concurrence bilinear form $\mathcal{C}_n(\ket{\psi},\ket{\phi})= 
\overline{\bra{\phi}\mho\ket{\psi}}$, satisfying \cite{BB:03} $C_n(\ket{\psi})=|\mathcal{C}_n(\ket{\psi},\ket{\psi})|$, we find
\begin{equation}
\begin{array}{lll}
\mathcal{C}_n\big(\; S_z\ket{\psi_g^{s_z}(h)},S_z\ket{\psi_g^{s_z}(h)} \;\big)
&=&s_z^2\; 
\mathcal{C}_n \big(\; \ket{\psi_g^{s_z}(h)},\ket{\psi_g^{s_z}(h)}\; \big)\\
\\
&=&\bra{\psi_g^{s_z}(h)} \; S_z^{\dagger}\mho S_z \mho^{-1}\mho\; \ket{\psi_g^{s_z}(h)}\\
\\
&=&\bra{\psi_g^{s_z}(h)}\; S_z^{\dagger}(-S_z)\mho\; \ket{\psi_g^{s_z}(h)}\\
\\
&=& -s_z^2\; \mathcal{C}_n\big(\; 
\ket{\psi_g^{s_z}(h)},\ket{\psi_g^{s_z}(h)}\; \big).\;
\end{array}
\end{equation}
Here we have used the reality of the eigenstates and the eigenvalues $s_z$, and the fact that $S_z$ is time-reversal antisymmetric.  Consequently, 
\begin{equation}
C_n\big( \; \ket{\psi_g^{s_z}(h)} \; \big)\ =\ 0 \qquad \mbox{if}\ s_z\neq 0.\;
\label{noconcsz}
\end{equation}

The $n$-concurrence of the ground state of $H_{XX}(0)$ is one as per 
Prop. \ref{prop:us}, therefore it is identified with the quantum number $s_z=0$.  As the magnetic field is increased, the energy gap decreases between the state $\ket{\psi_g^{0}(h)}$ and a state with different  symmetry, $\ket{\psi_g^{|s_z|=2}(h)}$, until they become degenerate at a magnetic field strength $h_{crit}$.  At this value of $h$, the ground state $n$-concurrence is zero.  For $|h|>h_{crit}$, it is energetically favorable for the spins to align meaning the spin projection $|s_z|$ can only increase so the $n$-concurrence remains zero by Eq. \ref{noconcsz}.  In order to identify the degeneracy point we must consider two cases:   the situtation with $n/2$ even, and $n/2$ odd.  For both we assume $J>0$, although a completely analogous argument can be made for $J<0$.        

\subsection{Case $n/2$ even}
At zero magnetic field and for $J>0$ the ground state energy is the lowest eigenvalue of $H^+$: 
\begin{equation}
E_0\ =\ 
-8J\sum_{k=1}^{n/4}\cos\bigg(\frac{\pi(2 k-1)}{n}\bigg)\ =\ -4J\csc(\pi/n)\;
\label{Eground}
\end{equation}
corresponding to the ground state 
\begin{equation}
\ket{\psi_g^0(|h|<h_{crit})}=\prod_{k=n/4+1}^{n/2}A_{2k-1}^{\dagger}A_{-(2k-1)}^{\dagger}\ket{0}.\;
\end{equation}
As $h$ increases from zero, $E_0$ becomes degenerate with the lowest eigenvalue of $H^-$: 
\begin{equation}
E_{min}^-=-8J\sum_{k=1}^{n/4}\cos\bigg(\frac{\pi 2 k}{n}\bigg)-2h-4J=-4J\cot(\pi/n)-2h\;
\end{equation}
corresponding to the state
\begin{equation}
\ket{\psi_g^{-2}(h_{crit})}=\bigg[\prod_{k=n/4-1}^{n/2-1}A_{2k}^{\dagger}A_{-2k}^{\dagger}\bigg]A_n^{\dagger}\ket{0}.\;
\end{equation}
The magnetic field strength where this degeneracy occurs is
\begin{equation}
h_{crit}\; =\; -2J[\cot(\pi/n)-\csc(\pi/n)]
\; =\; 2J\tan\bigg(\frac{\pi}{2n}\bigg).\;
\end{equation}

\subsection{Case $n/2$ odd}
At zero magnetic field and for $J>0$ the ground state energy is the lowest eigenvalue of $H^-$: 
\begin{equation}
E_0=-8J\sum_{k=1}^{(n-2)/4}\cos\bigg(\frac{\pi2 k}{n}\bigg)=-4J\csc(\pi/n)\;
\end{equation}
corresponding to the ground state 
\begin{equation}
\ket{\psi_g^{0}(|h|<h_{crit})}=\bigg[\prod_{k=(n+2)/4}^{n/2-1}A_{2k}^{\dagger}A_{-2k}^{\dagger}\bigg]A_n^{\dagger}\ket{0}.\;
\end{equation}
As $h$ increases from zero, $E_0$ becomes degenerate with the lowest eigenvalue of $H^+$: 
\begin{equation}
E_{min}^+=-8J\sum_{k=1}^{(n-2)/4}\cos\bigg(\frac{\pi (2 k-1)}{n}\bigg)-2h=-4J\cot(\pi/n)-2h\;
\end{equation}
corresponding to the state
\begin{equation}
\ket{\psi_g^{-2}(h_{crit})}=\prod_{k=n/4+1}^{n/2}A_{2k-1}^{\dagger}A_{-(2k-1)}^{\dagger}\ket{0}.\;
\end{equation}
The magnetic field strength where this degeneracy occurs is again
\begin{equation}
h_{crit}=2J\tan\bigg(\frac{\pi}{2n}\bigg).\;
\end{equation}
    
\section{Thermal entanglement}
\label{sec:therm}
The thermal state of a system with intrinsic Hamiltonian $H$ is given by
\begin{equation}
\rho_T\ =\ \frac{e^{-\beta H}}{Z}\ =\ \sum_j\sum_{k=1}^{g_j} \frac{e^{-\beta E_j}\ket{\chi_j^k}\bra{\chi_j^k}}{Z},\;
\end{equation}
where $\beta=1/k_BT$, with $k_B$ the Boltzmann constant, and the $\{\ket{\chi_j^k}\}$ are the energy eigenstates of $H$ corresponding to energy $E_j$ with degeneracy $g_j$.  Here $Z=\mbox{Tr}[e^{-\beta H}]$ is the partition function.  

In the case that the intrinsic Hamiltonian is time reversal symmetric, the spin-flipped thermal state satisfies
\begin{equation}
\begin{array}{lll}
\mho\rho_T\mho^{-1}&=&\mho e^{-\beta H}\mho^{-1}/Z\\&=&e^{-\beta\mho H \mho^{-1}}/Z\\&=&\rho_T.\;
\end{array}
\end{equation}
Thus $\{\lambda_j\}=\mbox{spec}[(\rho_T\mho\rho_T\mho^{-1})^{1/2}]=\mbox{spec}[\rho_T]$ is just the set of probabilities to be in a thermal eigenstate.  Using Eq.\ref{concmixed} we then find
\begin{equation}
C_n(\rho_T)=\max\bigg\{0,2e^{-\beta E_0}/Z-1\bigg\},\;
\label{conc}
\end{equation}
where $E_0$ is the ground state energy of $H$.  Because the partition function is a sum of positive terms, it is confirmed that $C(\rho_T)=0$ if the ground state of $H$ is degenerate.

The Hamiltonian $H_{XX}(h)$ is not time reversal symmetric for $h\neq 0$.  However, the concurrence for thermal states with a finite magnetic field is simply proportional to the concurrence at zero field as we now show.  Using the notation $Z(h)=\mbox{Tr}[e^{-\beta H_{XX}(h)}]$, we find
\begin{equation} 
\begin{array}{lll}
(\rho_T\mho \rho_T\mho^{-1})^{1/2}&=&\bigg(\frac{e^{-\beta H_{XX}(h)}}{Z(h)}\mho \frac{e^{-\beta H_{XX}(h)}}{Z(h)}\mho^{-1}\bigg)^{1/2}\\
\\
&=&\bigg(\frac{e^{-\beta H_{XX}(h)}}{Z(h)}\frac{e^{-\beta \mho H_{XX}(h)\mho^{-1}}}{Z(h)}\bigg)^{1/2}\\
\\
&=&\bigg(\frac{e^{-\beta H_{XX}(h)}e^{-\beta H_{XX}(-h)}}{Z(h)^2}\bigg)^{1/2}\\
\\
&=&\bigg(\frac{e^{-2\beta H_{XX}(0)}}{Z(h)^2}\bigg)^{1/2}\\
\\
&=&\frac{Z(0)}{Z(h)}\rho_T|_{h=0},\;
\end{array}
\label{proportional}
\end{equation}
where $\rho_T|_{h=0}$ denotes the thermal ensemble at $h=0$.   In deriving Eq. \ref{proportional} we have used Eq. \ref{reversed} in the third line and the fact that $[H_{XX}(0),hS_z]=0$ in the fourth line.  The concurrence is
\begin{equation}
C_n(\rho_T)=\frac{Z(0)}{Z(h)}\max\bigg\{0,2e^{-\beta E_0}/Z(0)-1\bigg\}.\;
\label{concgen}
\end{equation}
where $E_0$ is the ground state energy of $H_{XX}(0)$.
\begin{figure}
\begin{center}
\includegraphics[scale=0.26]{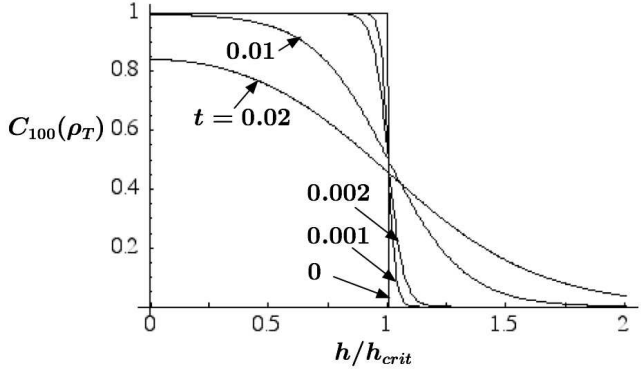}
\caption{\label{fig:2}Plot of the $n$-concurrence of thermal states in $100$ qubits of the Hamitonian $H_{XX}(h)$ as function of magnetic field strength $h$.  Shown are plots at several temperatures in units of the interaction strength: $t=k_BT/J$.  As the temperature drops to zero, the $n$-concurrence as a function of $h$ approaches a step function with a discontinuity at the critical magnetic field strength $h_{crit}$.  For $n=100$, $h_{crit}=2J\tan(\pi/100)\approx 0.0314J$.}
\end{center}
\end{figure}

\begin{figure}
\begin{center}
\includegraphics[scale=0.26]{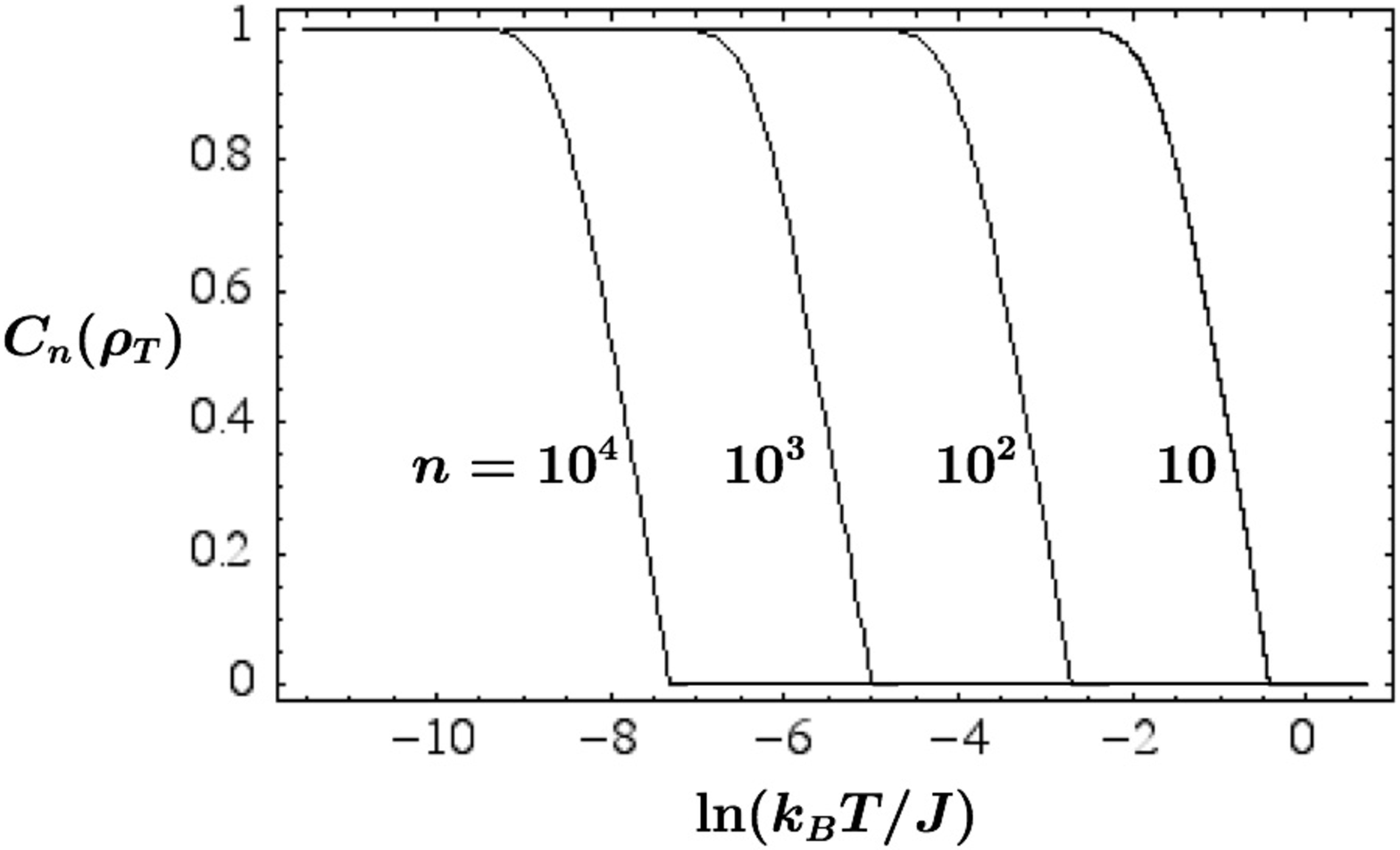}
\caption{\label{fig:3}Plot of the $n$-concurrence as a function of logarithm of the temperature for thermal states of the quantum XX model with zero magnetic field.  The concurrence is plotted for several values of the total number of spins $n$.}
\end{center}
\end{figure}

For the isotropic XX model, the partition function is calculated using the the exact diagonalization in Sec. \ref{sec:gse}  \cite{Katsura}
\begin{equation}
\begin{array}{lll}
Z(h)&=&2^{n-1}\big(\prod_{k=1}^{n/2} \cosh^2(\epsilon_{2k-1}\beta)+\prod_{k=1}^{n/2} \sinh^2(\epsilon_{2k-1}\beta)\\
& &+\prod_{k=1}^{n/2-1} \cosh^2(\epsilon_{2k}\beta)\cosh(\epsilon_0\beta)\cosh(\epsilon_n\beta)\\
& &-\prod_{k=1}^{n/2-1} \sinh^2(\epsilon_{2k}\beta)\sinh(\epsilon_0\beta)\sinh(\epsilon_n\beta)\big).\;
\end{array}
\label{part}
\end{equation}
Using this expression for $Z(h)$ and the value of the ground state energy in Eq. \ref{Eground}, the concurrence is readily computed, (see Figs. \ref{fig:2} and \ref{fig:3}.)

It is useful to know how mixed the quantum state can be before quantum correlations are lost.  For this purpose we define a threshold temperature $T_{th}$ as the temperature where the concurrence changes from a positive quantity from below $T_{th}$ to zero above $T_{th}$.  The value of $\beta_{th}=(k_BT_{th})^{-1}$ is obtained by solving the equation
\begin{equation}
2e^{-\beta_{th}E_0}/Z(0)-1\ =\ 0.\;
\label{sat}
\end{equation}

An approximate solution for $T_{th}$ can be found in the limit of large $n$.  The ground state energy Eq. \ref{Eground} per particle converges to,  
\begin{equation}
\lim_{n\to\infty}\frac{E_0}{n}\ =\ 
\lim_{n\to\infty}\frac{-4J\csc(\pi/n)}{n}\ =\ 
-\frac{4J}{\pi}\;
\end{equation}
In the same limit, $\epsilon_{2k-1}\approx \epsilon_{2k}$, and the logarithm of the partition function per particle converges to
\begin{equation}
\lim_{n\to\infty}\frac{\ln Z(0)}{n}=\frac{2}{\pi}\int_0^{\pi/2} d\omega \ln[2\cosh(2J\beta\cos(\omega))].\;
\end{equation} 
We can re\"express the integral as an infinite series in $(J\beta)^{-1}$:
\begin{equation}
\begin{array}{lll}
{\displaystyle \lim_{n\to\infty}\frac{\ln Z(0)}{n}}&=&4J\beta/\pi+\frac{2}{\pi}\int_0^{\pi/2}d\omega \ln[1+e^{-4J\beta \cos(\omega)}]\\
&=&4J\beta/\pi+n\sum_{k=1}^\infty \frac{2(-1)^{k+1}}{\pi k}\int_0^{\pi/2}d\omega e^{-4kJ\beta\cos(\omega)}\\
&=&4J\beta/\pi+\sum_{k=1}^\infty \frac{(-1)^{k+1}}{k}[{\bf I}_0(4kJ\beta)-{\bf L}_0(4kJ\beta)]\\
&\approx& 4J\beta/\pi-\sum_{k=1}^{\infty}\frac{(-1)^{k+1}}{\pi k}\sum_{m=0}^\infty \frac{(-1)^{m+1}\Gamma (1/2+m)}{\Gamma (1/2-m)(2kJ\beta)^{2 m+1}}\\
&=&4J\beta/\pi+\sum_{m=0}^\infty \frac{(-1)^{m} (1-2^{-2m-1})\zeta(2m+2)\Gamma (1/2+m)}{\pi\Gamma (1/2-m)(2J\beta)^{2 m+1}}\;
\end{array}
\end{equation}
where ${\bf I}_0(x)({\bf L}_0(x))$ is the zeroth order Bessel(Struve) function and $\zeta(x)$ is the Riemann Zeta function.  The approximation in the fourth line is valid when $J\beta\gg 1$ \cite{NBS}.  Inserting the two asymptotic expressions into Eq. \ref{sat}, and neglecting terms that fall off like $(J\beta)^{-3}$ and faster, we find
\begin{equation}
k_BT_{th}\approx\frac{J24\ln 2}{n\pi}\;
\label{Tcrit}
\end{equation}
One can then conclude that entanglement as measured by the $n$-concurrence is greater than zero when the temperature is less than the interaction energy per particle.
The derived expression for $T_{th}$ underestimates the exact value $T_{th}(\mbox{exact})$ by roughly $20\%$ (see Table \ref{tab:1}.)    

We compare our result with a related threshold temperature studied in \cite{Wang:02} for a ring of qubits coupled via the antiferromagnetic Heisenberg interaction.  There it was found that the threshold temperature for nearest neighbor $2$-concurrence approaches a constant in the thermodynamic limit.  This seems to indicate that quantum correlations between interacting qubit pairs are insensitive to changes in the character of the global thermal state as the system size increases.  In contrast, for the XX model, the threshold temperature for the $n$-concurrence decreases inversely with the number of qubits.  While we study a different spin chain Hamiltonian, qualitatively we expect a different threshold temperature for $n$-concurrence versus $2$-concurrence in any spin chain model.  This is because any  time reversal symmetric Hamiltonian, of which the XYZ model is one type, has a maximally concurrent ground state for any even $n$ provided it is non-degenerate.  The size dependence of $T_{th}$ is a consequence of the fact that the ground state population decreases with the size of the system for a fixed temperature.
      
\begin{table}
\caption{\label{tab:1}Ratio of the approximate analytic value of threshold temperature $T_{th}$ from Eq. \ref{Tcrit} to the numerically computed exact value $T_{th}^{p,o}$(\mbox{exact}) for periodic (p) and open (o) boundaries.  The ratio is computed for several decades of the system size $n$.}
\begin{ruledtabular}
\begin{tabular}{ccccc}
\hline
$n$ & $10$ & $10^2$ & $10^3$ & $10^4$ \\
\hline
$\frac{T_{th}}{T_{th}^p(\mbox{exact})}$ & $0.8159$ & $0.7960$ & $0.7959$ & $0.7959$ \\
\hline
$\frac{T_{th}}{T_{th}^o(\mbox{exact})}$ & $0.9979$ & $0.8969$ & $0.8887$ & $0.8879$ \\
\hline
\end{tabular}
\end{ruledtabular}
\end{table}

\section{Open boundaries}
\label{sec:openbndr}

The previous analysis was done for the quantum XY model in one dimension with periodic boundaries.  In most experimental situations it will be easier to construct the spin chains with open boundaries.  In this section we rederive the critical magnetic field and threshold temperature for this case.

The quantum XX model in one dimension with open boundaries is
\begin{equation}
\begin{array}{lll}
H_{XX}(h)&=&\sum_{j=1}^{n-1} J(\sigma_j^x\sigma^x_{j+1}+\sigma^y_j\sigma^y_{j+1})+h\sum_{j=1}^n\sigma_j^z\\
&=&\sum_{j=1}^{n-1} 2J( \sigma_j^+\sigma_{j+1}^-+\sigma_j^-\sigma_{j+1}^+)+h\sum_{j=1}^n\sigma_j^z.\;
\end{array}
\label{HXYfirst}
\end{equation}
It is assumed that $n$ is even.  After the transformation from Pauli operators to fermionic operators, the Hamiltonian assumes the form,
\begin{equation}
H_{XX}(h)=nh-2J\sum_{j=1}^{n-1}(a_{j+1}^{\dagger}a_j+a_j^{\dagger}a_{j+1})-2h\sum_{j=1}^n a_j^{\dagger}a_j\;
\end{equation}
The Hamiltonian is a quadratic form on the vector space of the creation and annihilation operators $\{a_j,a_j^{\dagger}\}$.  Written as an $n\times n$ matrix in the basis $\{a_1,a_2,\ldots a_n\}$, $H_{XX}(h)$ is a sum of two terms.  The first is proportional to the identity and the second is a tridiagonal matrix with elements $-2h$ on the diagonal and elements $-2J$ on the nearest off diagonal.  The eigenvalues and eigenvectors of the tridiagonal can be solved for exactly yielding 
$H_{XX}(h)=\sum_k\Lambda_kb_k^{\dagger}b_k+C$, where $\Lambda_k=-2h-4J\cos(\pi k/(n+1))$, $b_j^{\dagger},b_j$ are the quasi-particle creation and annihilation fermionic operators, and $C$ is a constant.  The value of the constant is determined by demanding that $\mbox{Tr}[H_{XX}(h)]=0$.  Summing over all possible particle occupations, each particle energy $\Lambda_k$ appears $\sum_{l=0}^{n-1}{n-1 \choose l}=2^{n-1}$ times and the constant satisfies $-2^{n-1}\sum_{k=1}^n\Lambda_k+2^nC=0$ or $C=nh$.  The eigenenergies can be expressed compactly as
\begin{equation}
E=\bigg\{\sum_{k=1}^n \Lambda_k; \ \Lambda_k\in\{-2h-4J\cos(\pi k/(n+1)),0\}\bigg\}+nh\;
\end{equation}
At zero magnetic field and for $J\geq 0$, the ground state energy is
\begin{equation}
\begin{array}{lll}
E_0&=&-4J\sum_{k=1}^{n/2}\cos(\pi k/(n+1))\\
&=&-4J \bigg(\cos\bigg(\frac{n\pi}{4(n+1)}\bigg)\csc\bigg(\frac{\pi}{2(n+1)}\bigg)\sin\bigg(\frac{(n+2)\pi}{4(n+1)}\bigg)-1\bigg).
\end{array}
\label{Eground}
\end{equation}
The corresponding ground state is an eigenstate of $S_z=\sum_{j=1}^n\sigma_j^z$ with eigenvalue $s_z=0$.  As the magnetic field is increased from zero, this eigenstate becomes degenerate with the $s_z=-2$ eigenstate at the critical value
\begin{equation}
h_{crit}=2J\sin(\pi/(2(n+1))).\;
\end{equation}
For $|h|\geq h_{crit}$, the $n$-concurrence of the ground state is equal to zero.  Notice for $n\gg1$, $h_{crit}\approx J\pi/n$ which is the same asymptotic as in the XX model with periodic boundaries.

For the $n$-concurrence of thermal states, we need an expression for the partition function $Z(h)=e^{-\beta H_{HH}(h)}$.  In the representation of the quasi-particles, the Hamiltonian can be written as a tensor product:
\begin{equation}
H_{XX}(h)=\bigotimes_{k=1}^n\begin{pmatrix}
    e^{-\beta (\Lambda_k+h) } &  0  \\
     0 &  e^{-\beta h}
\end{pmatrix}
\end{equation}
The partition function is then
\begin{equation}
\begin{array}{lll}
Z(h)&=&\prod_{k=1}^ne^{-\beta h}(e^{-\beta \Lambda_k}+1)\\
&=&e^{-\beta(nh+\sum_{k=1}^n\Lambda_k/2)}\prod_{k=1}^n2\cosh(\beta\Lambda_k/2)\\
&=&2^n\prod_{k=1}^n\cosh(\beta(h+2J\cos(\pi k/(n+1)))\;
\end{array}
\end{equation}
Given an expression for the partition function, the concurrence can be calculated for any magnetic field and temperature using Eq. \ref{concgen}.  The threshold temperature can be computed by the same proceedure used in Sec. \ref{sec:therm}.  In the large $n$ limit, the ground state energy at zero magnetic field is
\begin{equation}
\lim_{n\to\infty}\frac{E_0}{n}\ =\ 
-\frac{4J}{\pi},\;
\end{equation}
which is the same asymptotic as the case with periodic boundaries.  Again in the large $n$ limit we find
\begin{equation}
\lim_{n\to\infty}\frac{\ln Z(0)}{n}=\frac{2}{\pi}\int_0^{\pi/2} d\omega \ln[2\cosh(2J\beta\cos(\omega))].\;
\end{equation} 
and the threshold temperature is therefore
\begin{equation}
k_BT_{th}\approx\frac{J24\ln 2}{n\pi}.\;
\label{Tcrit}
\end{equation}
In Table \ref{tab:1} the exact value of the threshold temperature is computed numerically for comparison with the open boundaries case.
\section{Measurement}
\label{sec:meas}

For any measure of entanglement in a many body system it is important to find a practical method to construct its corresponding observable or set of observables.  Computing the $n$-concurrence on pure states is generically hard because the time reversal operator is not physical and therefore does not correspond to any single observable on the system.  It has been shown that for pure states the $n$-tangle, which is the square of the $n$-concurrence, can be computed in terms of the multiphoton Stokes scalar \cite{Teich:03}.  There are an exponentially large number of multiphoton Stokes parameters that need to be measured in order to compute this scalar.  Nevertheless, it may still be more efficient than resorting to direct state tomography over all $n$ qubits which generically requires $4^n$ measurements.  

A notable feature of Hamiltonians over spin chains such as the XX model is the reality of the Hamiltonian.  This implies that the energy eigenstates are real and the concurrence of the ground state is then
\begin{equation}
C_n(\ket{\psi_g})=|\bra{\psi_g}\mho\ket{\psi_g}|=|\bra{\psi_g}\prod_{j=1}^n(-i\sigma^y_j)\ket{\psi_g}|\;
\end{equation}

Measuring the expectation value of the many body operator $S=\prod_{j=1}^n(-i\sigma^y_j)$ can be done in various ways and the preferred technique will depend on the particular system.  One approach would be to introduce an ancillary qubit prepared in the state $\ket{+_x}_A=1/\sqrt{2}(\ket{0}_A+\ket{1}_A)$ and use this ancilla as a control in a sequence of $n$ controlled rotation gates $\prod_{j=1}^n \Lambda_{A,j}(-i\sigma^y)$.  After measuring the ancilla in the $\ket{\pm_x}_A$ basis, the concurrence is obtained from the measurement probabilities as
$|\mbox{Prob}(A==1)-\mbox{Prob}(A==0)|\; =\; C_n(\ket{\psi_g})$. 


Measuring the $n$-concurrence for mixed states is much more difficult and will probably require some amount of state tomography.  In the case of thermal samples, if the temperature can be measured by some means then the results in this paper place a bound on how high it can be before quantum correlations as measured by $C_n$ are lost.  The analysis above shows that nearly maximal $n$-concurrence is achievable in the XX model with zero magnetic field for sufficiently small temperatures.  Because the temperature must be less than the interaction energy per particle this will be experimentally challenging to observe.  

One candidate system may be cold trapped neutral atoms.  Recently, it was proposed to simulate the quantum XY model in a lattice of spins trapped in an optical lattice \cite{Duan:03}.  The lattice can be designed to trap an antiferromagnetic array of spin polarized atoms and the interactions can be engineered via ``always on" ground state collisions between nearest neighbor atoms in the lattice.  The advantage of using trapped atoms is that the lattice can be loaded from an atomic Bose Einstein Condensate (BEC) which are routinely prepared at extremely low temperatures ($\sim 50$ nK).   For example, in a 1D lattice with 100 qubits at a temperature of $50$ nK, one would need interaction strengths of $J\approx\hbar2\pi\times 16 \mbox{KHz}$ in order to be in the regime of non-zero $n$-concurrence and an interaction strength $J=\hbar 2\pi\times 115\mbox{KHz}$ to be close to maximal $n$-concurrence.  Generating a sufficiently strong interaction between neutral atoms is challenging, however, for appropriately tuned trapped potentials the collisional interactions can be on the order of a few tens of kilohertz.  


\section{Conclusions}
\label{sec:concl}
We have investigated the behavior of many qubit entanglement in one dimensional spin chains.  The entanglement measure used, the $n$-concurrence, is a global measure in the sense that its value goes to zero if any qubit is completely disentangled.  This measure is appropriate in the context of spin chains because it probes the onset of a break in the time reversal symmetry when a magnetic field is introduced.  For pure states in $n$ even qubits, the $n$-concurrence is maximal at zero magnetic field.  In the XX model, the entanglement is not immediately lost when a magnetic field perturbation is added, rather it jumps abruptly to zero at a critical magnetic field strength corresponding to the first degeneracy point.  Using the symmetry properties of the Hamiltonian under conjugation by the time reversal operation we are able to obtain expressions for the entanglement of thermal systems of arbitrary size $n$.  The entanglement was shown to be finite below a certain critical magnetic field strength and threshold temperature.
 
Several outstanding issues remain.  First, the objection may be raised that maximal $n$-concurrence in the ground states of $H_{XY}(0)$ does not imply global entanglement whatsoever because subglobal entanglement, i.e. tensor products of time reversal symmetric states, also have maximal $C_n$ (see Sec. \ref{sec:prop}).  However, it has been proven \cite{Nielsen,Osterloh} that the ground state of the XX model has non vanishing $2$-concurrence over all pair separations.  Hence, there can be no way to partition the ground state into disentangled subsets of qubits and therefore the entanglement is truly global.    Second, it is of interest to study how inhomogeneous couplings affect the $n$-concurrence.  Perhaps the behavior of global entanglement could give a signature of local defects.  Finally, because time reversal is not a spatial symmetry, it may be possible to investigate entanglement in the usually more complicated model of spin chains embedded in $d>1$ dimensions.

Another question for future research is whether the highly entangled 
ground state of a spin chain can be efficiently transformed into a 
state useful from the point of view of quantum information processing 
({\tt QIP}).
Spin chains are a promising architecture to implement many 
{\tt QIP} tasks such as entanglement distribution \cite{Brennen:03}, quantum 
state swapping \cite{Khaneja:02}, and quantum 
computation \cite{Benjamin:03}.  Preparing and observing many-body 
entanglement will be an important first step to realizing these 
challenging tasks.  It is appealing to try to generate useful entanglement 
by allowing the system to naturally cool to its ground state.  The results 
obtained in this paper help define limits to the environmental 
conditions under which one type of entanglement (the $n$-concurrence) 
can be prepared in this way.  It is known \cite{BB:03}
that all states of fixed $n$-concurrence are orbits of the Lie (symmetry)
group $K\cong SO(2^n)$ of all unitary evolutions which admit time
reversal antisymmetric Hamiltonains.  
In particular, there is some (and there are many) Hamiltonian(s) $H$ with
$H \; = \; - \mho H \mho^{-1}$ and
\mbox{$\big| \mbox{GHZ} \big\rangle \ = \ k \ket{\psi_g}$}
for $k \; = \; \mbox{exp}(\; -i Ht/\hbar \; )$.
It is unclear whether 
there exist \emph{efficient} operations in 
$K$ to transform from one to the other.

\appendix 

\section{Concurrence of mixed states}
\label{app}

This appendix expands upon published work of Wootters and Uhlmann
\cite{Wootters:98,Uhlmann:00},
recalling their closed-form expression for the minimum Eq. 
\ref{concmixed}.  
It is, to our 
knowledge, the first treatment of this problem that is complete 
and self-contained.

\subsection{Notation and conventions}

For either scalars or vectors, we denote complex conjugates by an
overline.  Also, we often use both real and complex operators.
Thus, we forego bra-ket notation in favor of the set 
$\{e_j \}_{j=0}^{\ell-1}$ of
\emph{standard basis vectors}, i.e. column matrices
with $(e_j)_{\ell} = \delta_j^\ell$.  The symbols $\mathbb{R}^{p \times q}$
and $\mathbb{C}^{p \times q}$ denote $p \times q$ real, complex matrices
respectively.  Finally, we use a dagger for adjoint throughout.  For example,
$e_j^\dagger$ is formally equivalent to a bra if the operator is intended
to be complex.

\subsection{Background}

Throughout, we distinguish $\mathbb{C}^\ell$ and 
$\mathbb{R}^{2\ell}$ as vector spaces over $\mathbb{R}$.  
Thus, \emph{throughout}
we use following choice of $\mathbb{R}$-isomorphism:
\begin{equation}
{x} = {a} + i {b} \; \in \mathbb{C}^\ell \ \longleftrightarrow
\ {x}_{\mathbb{R}} = \left( \begin{array}{r} a \\ b \\ \end{array}
\right) \; \in \mathbb{R}^{2\ell}
\end{equation}
If $w=u+iv \in \mathbb{C}^{\ell \times \ell}$ is
a $\mathbb{C}$-linear map, we note the corresponding $\mathbb{R}$-linear
map under the isomorphism.  Specifically,
$w(a+ib) = (u+iv)(a+ib) = (ua-vb)+i(va+ub)$ for
\begin{equation}
w = u+iv \; \in \mathbb{C}^{\ell \times \ell} 
\longleftrightarrow
w_{\mathbb{R}} = \left( \begin{array}{rr}
u & -v \\ v & u \\ \end{array} \right) \; \in \mathbb{R}^{2\ell \times 2\ell}
\end{equation}
Finally, we note 
that ${x} {x}^\dagger$ is \emph{not} 
$x_{\mathbb{R}} x_{\mathbb{R}}^T$. 
Rather, by $\rho_\mathbb{R}$ we intend $\mathbb{R}$-linear
map that extends the $\mathbb{C}$-linear map given by $\rho$.  Computing,
as $x=a+ib \in \mathbb{C}^\ell$, we see
$\rho = x x^\dagger = (a+ib)(a^T - ib^T) = (aa^T + b b^T) +
i (b a^T - a b^T)$, for
\begin{equation}
\begin{array}{ll}
\rho = x x^\dagger, x = a+ib, & \rho_{\mathbb{R}} = 
\left( \begin{array}{rr}
aa^T + b b^T & a b^T - b a^T \\
b a^T - a b^T & a a^T + b b^T \\
\end{array}
\right)
\\
\rho = u + i v \in \mathbb{C}^{\ell \times \ell}, & 
\rho_{\mathbb{R}} = \left( \begin{array}{rr}
u & -v \\ v & u \\ \end{array} \right)
\end{array}
\end{equation}

We also should breifly comment on the structure of $u_{\mathbb{R}}$
for $u u^\dagger = I_\ell$, $u \in \mathbb{C}^{\ell \times \ell}$.  Clearly any
unitary map of $\mathbb{C}^{\ell}$ will lift to an orthogonal map
of $\mathbb{R}^{2\ell}$, but this is not the complete structure.
Complex multiplication by $i$ lifts as a matrix
$J = \left( \begin{array}{rr} {\bf 0} & - {I} \\ { I} & {\bf 0} \\
\end{array} \right)$.  Then an orthogonal map $o$ of $\mathbb{R}^{2\ell}$
has $o=u_{\mathbb{R}}$ for some unitary $u$ iff $o J = J o$.
Equivalently, a unitary map is an orthogonal map which is also
$\mathbb{C}$-linear.  Adding a tad more language, as one speaks
of the unitary group $U(\ell)$ there is also a real symplectic group
\cite[pg. 446]{Helgason:01} given by
\begin{equation}
Sp(\ell,\mathbb{R}) = \{ \ A \in 
\mathbb{R}^{2\ell \times 2\ell} \; ; \; A^T J A = J \ \}
\end{equation}
Then it is standard that $Sp(\ell,\mathbb{R}) \cap O(2\ell) \cong U(\ell)$
\cite[pg. 447, Lemma2.1(c) of Chap.X.2]{Helgason:01}
as we have verified above.

We also note generalities on time-reversal
symmetry operators, denoted $\Theta$, on a Hilbert state-space
of complex dimension $N$.  Such a $\Theta$
is an \emph{$\mathbb{R}$-linear map} of the state space possessing
(i) complex anti-linearity, (ii) orthogonality $\Theta^T \Theta=I_{2N}$,
and (iii) projectively involutivity 
$\Theta^2= (\mbox{e}^{i \varphi} I_{N})_{\mathbb{R}}$, 
$\phi \neq 0$.  Given that $\tau$ is the
$\mathbb{R}$-linear map corresponding to complex conjugation on
$\mathbb{C}^N$, we note that any antilinear map (hence $\Theta$) may
be written as $\Theta=\omega_{\mathbb{R}} \tau$ for $\omega \in
\mathbb{C}^{N \times N}$.  Due to orthogonality of $\Theta$ and
$\tau$, $\omega_{\mathbb{R}}$ is orthogonal.  Hence $\omega$ is unitary,
given $\mathbb{C}$-linearity.  We further claim that $\mbox{e}^{i \varphi}$
is $\pm 1$.  Indeed, $\Theta^2 = \omega_{\mathbb{R}} \tau \omega_{\mathbb{R}}
\tau = \omega_{\mathbb{R}} \overline{\omega}_{\mathbb{R}}$, so that
$\omega \overline{\omega} = \mbox{e}^{i \varphi} I_N$.  But
$\mbox{Tr}(\omega \overline{\omega}) = \mbox{Tr}(\overline{\omega}\omega)$,
demanding that the trace is real.  Hence $\mbox{e}^{i \varphi}$ is
$\pm 1$.
We refer the $+1$ case as \emph{bosonic} time-reversal symmetry operators,
and the $-1$ case are \emph{fermionic} time-reversal symmetry operators.

\subsection{Analysis of
$\bigg| \sum_{j=0}^{\ell-1} \mbox{e}^{i \theta_j} \lambda_j \bigg|$}

The following proposition will be quite important to what
follows.  Thus, we present a careful argument.  It improves an
earlier inductive proof of the second author and was suggested by
Dianne O'Leary.

\vbox{
\begin{proposition}
\label{prop:bigmess}
Let $\lambda_0 \geq \lambda_1 \geq \lambda_2 \geq
\cdots \geq \lambda_{\ell-1} \geq 0$ be an ordered set
of real numbers.
\begin{itemize}
\item  If $\sum_{j=1}^{\ell-1} \lambda_j < \lambda_0$, then
\begin{equation}
\lambda_0 - \sum_{j=1}^{\ell-1} \lambda_j = 
\mbox{min} \bigg\{ \ \bigg| \sum_{j=0}^{\ell-1} \mbox{e}^{i \theta_j}
\lambda_j \bigg| \ ; \ 
\theta_0, \cdots \theta_{\ell-1} \in \mathbb{R} \ \bigg\}
\end{equation}
\item  If $\sum_{j=1}^{\ell-1} \lambda_j \geq \lambda_0$, then there exist
$\{ \theta_j \}_{j=0}^{\ell-1}$ so that
\begin{equation}
\sum_{j=0}^{\ell-1} \mbox{e}^{i \theta_j} \lambda_j = 0
\end{equation}
\end{itemize}
\end{proposition}
}

The proof uses the followig two lemmas.  The first is
a standard result in combinatorics.

\begin{lemma}
\label{lem:partition}
Let $\lambda_0 \geq \lambda_1 \geq \lambda_2 \geq \cdots \geq \lambda_{\ell-1}
\geq 0$, and label $S = \{0,1,\cdots,\ell-1\}$.  Then there exists
a partition $S = S_1 \sqcup S_2$ such that
\begin{equation}
\lambda_0 \ \geq \  
\sum_{j \in S_1} \lambda_j - \sum_{j \in S_2} \lambda_j \ \geq \ 0
\end{equation}
\end{lemma}

\begin{proof}
Let $S=S_1 \sqcup S_2$ be a partioning chosen to minimize
$\mbox{max} 
\bigg\{ \sum_{j \in S_1} \lambda_j - \sum_{j \in S_2} \lambda_j, 0\bigg\}$.
Label $\delta \geq 0$ to be this minimal difference, i.e.
$\delta = \sum_{j \in S_1} \lambda_j - \sum_{j \in S_2} \lambda_j$.

Assume \emph{by way of contradiction} that $\delta > \lambda_0$.
In particular, some element of $S_1$ must be nonzero, say $\lambda_k>0$.
Now place $\tilde{S}_1=S_1-\{\lambda_k\}$ and 
$\tilde{S}_2=S_2 \sqcup \{\lambda_k\}$.
Then label $\tilde{\delta} \geq 0 $ by
\begin{equation}
\tilde{\delta} \ = \ 
\bigg|
\sum_{j \in \tilde{S}_1} \lambda_j - \sum_{j \in \tilde{S}_2} \lambda_j \bigg|
\ = \
| \delta - 2 \lambda_k |
\end{equation}
Now either $|\delta - 2 \lambda_k| = \delta - 2 \lambda_k$ or else
$|\delta - 2 \lambda_k| = 2\lambda_k - \delta$.  In the former case,
note that $\delta - \tilde{\delta} =
2\lambda_k > 0$, i.e. $\delta > \tilde{\delta}$ contradiction.
In the latter case, $\delta - \tilde{\delta} =
2\delta - 2\lambda_k \geq 2(\delta - \lambda_0) > 0$, contradiction.
\end{proof}

\begin{lemma}
\label{lem:theta}
Let $t_0 \geq t_1 \geq 0$, and label 
$L(\theta) = | t_0 + \mbox{e}^{i \theta} t_1|$.  Then for every
$s \in [t_0-t_1, t_0+t_1]$, there is some $\theta_0 \in [0,\pi]$
such that $L(\theta_0)= s$.
\end{lemma}

\begin{proof}
Clearly $L(\theta)$ is continuous.  Now $L(\pi)=t_0-t_1$ and
$L(0)=t_0+t_1$, so that the result follows by the
Intermediate Value Theorem.
\end{proof}

\begin{proof}[\ of Prop. \ref{prop:bigmess}]
We begin with the first item.  Any $\{ \theta_j \}_{j=1}^{\ell-1}$ must have
$\bigg| \sum_{j=1}^{\ell-1} \mbox{e}^{i \theta_j} \lambda_j \bigg| \leq
\sum_{j=1}^{\ell-1} | \mbox{e}^{i \theta_j} \lambda_j |$.  Thus
\begin{equation}
\label{eq:clear}
\bigg| \sum_{j=0}^{\ell-1} \mbox{e}^{i\theta_j} \lambda_j \bigg|
\ \geq \ 
| \mbox{e}^{i \theta_0} \lambda_0| - 
\bigg| \sum_{j=1}^{\ell-1} \mbox{e}^{i \theta_j} \lambda_j \bigg|
\ \geq \ \lambda_0 - \sum_{j=1}^{\ell-1} \lambda_j
\end{equation}
On the other hand, $\sum_{j=1}^{\ell-1} \lambda_j < \lambda_0$ demands
the last expression of Equation \ref{eq:clear} is contained
within the set being minimized.

For the second part, label the \emph{($\ell-1$)}-index set
$S=\{1,\cdots,\ell-1\}$.
By Lemma \ref{lem:partition}, we may
partition $S= S_1 \sqcup S_2$ so that
\begin{equation}
\lambda_0 \ \geq \ \lambda_1 \ \geq \   
\sum_{j \in S_1} \lambda_j - \sum_{j \in S_2} \lambda_j \ \geq \ 0
\end{equation}
Label $t_0 = \sum_{j \in S_1} \lambda_j$ and
$t_1 = \sum_{j \in S_2} \lambda_j$.  Now
$t_0 + t_1 = \sum_{j=1}^{\ell-1} \lambda_j 
\geq \lambda_0$.  Hence by Lemma \ref{lem:theta},
there is some $\theta$ so that
\begin{equation}
| t_0 + \mbox{e}^{i \theta} t_1 | \ = \ 
\bigg| \sum_{j \in S_1} \lambda_j + \sum_{j \in S_2} \mbox{e}^{i \theta}
\lambda_j \bigg|  \ = \ \lambda_0
\end{equation}
Hence for some $\mbox{e}^{i \psi}$, we have
$\mbox{e}^{i \psi} \lambda_0 = \sum_{j \in S_1} \lambda_j + 
\sum_{j \in S_2} \mbox{e}^{i \theta} \lambda_j$.  Thus
$\mbox{e}^{i \psi} \lambda_0 + \sum_{j \in S_1} \mbox{e}^{i \pi} \lambda_j
+ \sum_{j \in S_2} \mbox{e}^{i(\theta + \pi)} \lambda_j = 0$.
\end{proof}

\subsection{Concurrence of Ensembles}

We set further notations regarding density matrices.  These make
later arguments more clear and terse.

\begin{appdefinition}
Let $\rho \in \mathbb{C}^{N \times N}$ be 
a density matrix,
i.e. $\rho = \rho^\dagger$, $\rho \geq 0$, and $\mbox{Tr}(\rho)=1$.  
A \emph{subnormalized
ensemble for $\rho$ of length $\ell$} is any collection of vectors 
$\mathcal{E}=\{ x_j \}_{j=0}^{\ell-1} \subset \mathbb{C}^N$
satisfying $\rho = \sum_{j=0}^{\ell-1} x_j x_j^\dagger$.
This is associated to a \emph{normalized ensemble of length $\ell$}
given by $\{ y_j \}_{j=0}^{\ell-1}$, where $y_j=\vec{0}$ if $x_j=\vec{0}$
and $y_j=x_j/|x_j|$ else.
We further define a right action
of the unitary group $U(\ell)$ on the set of subnormalized 
ensembles of length $\ell$ as follows.  If $u = (u_{jk}) 
\in \mathbb{C}^{\ell \times \ell}$, $u u^\dagger = I_\ell$, then
$\mathcal{E} \cdot u = \{ w_j \}_{j=0}^{\ell-1}$ where
$w_j = 
\sum_{k=0}^{\ell-1} x_k e_j^\dagger u e_k = \sum_{k=0}^{\ell-1} x_k u_{jk}$. 
\end{appdefinition}

\begin{appremark}
As always, we could produce a left action by considering the right
action of the adjoint.  However, we find that confusing in this context.
Checking the right action, consider $\{ w_j\}_{j=0}^{\ell-1}
= (\mathcal{E} \cdot u^1) \cdot u^2$ for $u^1, u^2$ unitary.  Then
\begin{equation}
w_j \ = \ \sum_{k=0}^{\ell-1} \sum_{p=0}^{\ell-1} x_p u_{jk}^1 u_{kl}^2
\ = \ \sum_{p=0}^{\ell-1} x_p \bigg( \; \sum_{k=0}^{N-1} u_{jk}^1 u_{kl}^2
\; \bigg)
\end{equation}
This is $\mathcal{E} \cdot (u^1 u^2)$.  We also remark that if
$\mathcal{E}$ is an ensemble for $\rho$, then so likewise is
$\mathcal{E} \cdot u$.  Indeed,
\begin{equation}
\begin{array}{l}
\sum_{j=0}^{\ell-1} w_j w_j^\dagger \ = \ 
\sum_{j=0}^{\ell-1} \sum_{k=0}^{\ell-1} \sum_{p=0}^{\ell-1}
x_k e_j^\dagger u e_k e_p^\dagger u^\dagger e_j x_p^\dagger 
\ = \\
\sum_{j=0}^{\ell-1} \sum_{k=0}^{\ell-1} 
x_k e_j^\dagger u u^\dagger e_j x_k^\dagger
\ = \ \sum_{k=0}^{\ell-1} x_k x_k^\dagger \ = \ \rho
\end{array}
\end{equation}
Thus, the $U(\ell)$ action respects the density matrix structure.
It has been proven \cite{HughstonEtAl:93}
the action is transitive on the set
returning $\rho$; \emph{any subnormalized ensemble for $\rho$ of length
$\ell$ arises in this way.}  Every $\rho$ possesses an ensemble
of length $n$, due to the spectral theorem.  We also remark that since
$\mbox{dim}_{\mathbb{C}} \mbox{End}_{\mathbb{C}}(\mathbb{C}^N)=N^2$, it is in
some sense wasteful to take $\ell > N^2$.  However, the arguments
would not simplify with this convention. Finally, note that evidently
any ensemble for $\rho$ must have length at least $\mbox{rank}(\rho)$.
\end{appremark}

\vbox{
\begin{appdefinition}
Let $\mathcal{E}=\{x_j\}_{j=0}^{N-1}$ be any ensemble of a 
fixed density matrix $\rho$.  The \emph{preconcurrence}
\hbox{$c_{2p}: \mathbb{C}^N \rightarrow \mathbb{C}$}
is
\begin{equation}
c_{2p}( x) \ = \ 
\left\{
\begin{array}{rr}
x^\dagger \Theta x = x^\dagger \omega \bar{x}, & x \neq \vec{0} \\
0, & x = \vec{0} \\
\end{array}
\right.
\end{equation}
Note that for $x \neq \vec{0}$, 
$|c_{2p}(x)|\; =\; |x|^2C_{2p}(x/|x|)$.
We then define the concurrence of an ensemble $\mathcal{E}$ to be
the following sum, for $\mathcal{E}=\{x_k\}_{k=0}^{\ell-1}$:
\begin{equation}
C_{2p}(\mathcal{E}) \ = \ \sum_{k=0}^{\ell-1} |c_{2p}(x_k)|
\ = \ \sum_{x \neq \vec{0} \in \mathcal{E}} |x|^2 C_{2p}(\; x/|x|\; ) 
\end{equation}
We also make the following definition, 
fixing some preferred ensemble $\mathcal{E}_0$ for $\rho$.
\begin{equation}
\begin{array}{lll}
C_{2p}^\ell(\rho)&=&\mbox{min}\{C_{2p}(\mathcal{E}) \; ;
\mathcal{E} \mbox{ is an ensemble of length } \ell 
\mbox{ for } \rho\} \ \\
&=& \mbox{min} \{ \; C_{2p}(\mathcal{E}_0 \cdot u) \; ; \; u \in U(\ell) \; \}
\end{array}
\end{equation}
We then define $C_{2p}(\rho) = 
\mbox{min} \{ \; C_{2p}^\ell(\rho) \; ; \; \ell \geq 1 \; \}$.
\end{appdefinition}
}


\vbox{
\begin{theorem}[\cite{Uhlmann:00}]
\label{thm:uhlmann}
View a given
time-reversal symmetry operator
$\Theta \in \mathbb{R}^{2N \times 2N} 
\cong \mbox{End}_{\mathbb{R}}(\mathcal{H}_n)$, 
and denote the $\mathbb{R}$-linear 
map corresponding to $\rho$ as $\rho_{\mathbb{R}} \in
\mathbb{R}^{2N \times 2N}$.  Since $\Theta$ is antiunitary,
in particular orthogonal, $\Theta^T=\Theta^{-1}$.
We write $\Theta = \omega_{\mathbb{R}} \tau$ for
$\omega \in \mathbb{C}^{N \times N}$ and $\tau$ the complex-conjugation
viewed within $\mbox{End}_{\mathbb{R}}(\mathbb{C}^N)$.
Moreover say $\{ \lambda_j \}_{j=0}^{N-1}=\mbox{spec}[M(\rho)]$ are the (real)
eigenvalues of 
\begin{equation}
\label{eq:Mrho}
\begin{array}{l}
M(\rho) = (\sqrt{\rho}\; \omega \; 
\overline{\rho} \; \omega^\dagger \; \sqrt{\rho})^{1/2},
\mbox{ noting that } \\
{[} \ (\sqrt{\rho}\; \omega \; 
\overline{\rho} \; \omega^\dagger \; \sqrt{\rho})^{1/2}\ {]}_{\mathbb{R}} = 
(\sqrt{\rho_{\mathbb{R}}} 
\Theta \rho_{\mathbb{R}} \Theta^T \sqrt{\rho_{\mathbb{R}}})^{1/2} \\
\end{array}
\end{equation}
ordered so that
$\lambda_0 \geq \lambda_1 \geq \lambda_2 \cdots \geq \lambda_{N-1} \geq 0$.
Then if $N=2^n$ and $\Theta$ bosonic, we must have
\begin{equation}
C_{2p}(\rho) \ = \ \mbox{max} 
\bigg\{ \ 0, \lambda_0 - \sum_{j=1}^{N-1} \lambda_j \ \bigg\}
\end{equation}
\end{theorem}
}

In view of Proposition \ref{prop:bigmess}, we prove 
this theorem in two steps.  Put
$T= \mbox{max} \{ 0, \lambda_0 - \sum_{j=1}^{N-1} \lambda_j \}$.
In the first step, we 
produce an ensemble $\mathcal{E}_{\mbox{min}}$ for $\rho$ 
(of length $N$) such that
the $T= C_{2p}(\mathcal{E}_{\mbox{min}})$.  In the second step,
we restrict to case $T \neq 0$ and 
show that any ensemble $\mathcal{E}$ for $\rho$ has
$C_{2p}(\mathcal{E}) \geq T$.  Each
of the two steps is organized within a subsection,
culiminating in Propositions \ref{prop:min_pos},
\ref{prop:min_zero}, and \ref{prop:lower_bound}.

\subsection{Existence of a Minimizing Ensemble}

\begin{lemma}
\label{lem:real_trick}
Let $A \in \mathbb{R}^{\ell \times \ell}$.  Then there is some
$o \in \mathbb{R}^{\ell \times \ell}$, $o o^T = I_\ell$, so that
$(o A o^T)_{00}=(oAo^T)_{11}=\cdots=(oAo^T)_{\ell - 1 \; \ell-1}$.
\end{lemma}

\begin{sketch}
Let $T= \frac{1}{\ell}\sum_{j=0}^{\ell-1} a_{jj}$, the average
value of the diagonal elements of $A$.  Since the trace of $o A o^T$
coincides with the original trace, we seek to set all diagonal elements
to be $T$.  If every diagonal element is $T$ already, then take $o=I_\ell$.

Else some diagonal element exceeds $T$, and some is less than $T$.
By choosing an
appropriate permutation matrix
$\Pi$ and relabeling $\Pi A \Pi^T$ as $A$, we may suppose without
loss of generality (WLOG)
that $a_{00}< T$, $a_{11}>T$.  Now label $R(t)= 
\mbox{e}^{-i t \sigma^y} \oplus I_{\ell -2}$.  Consider the continuous
function $t \mapsto [R(t) A R(t)^T]_{00}
=\cos^2(t)a_{00}-\sin^2(t)a_{11}$.
The Intermediate Value Theorem shows that
for some $t_0$ the diagonal entry is $T$.  Now induct on the number
of entries equal to $T$.
\end{sketch}


\begin{lemma}  
\label{lem:not_diag}
Let $\eta \in \mathbb{C}^{\ell \times \ell}$ be a
\emph{complex, perhaps non-Hermitian matrix}:  $\eta = \eta^T$.
Then there exists a unitary matrix $u \in U(\ell)$ so that
\begin{equation}
\label{eq:NOT_diagonalization}
u \; \eta \; u^T \ = \ \Lambda, \quad
\Lambda = \mbox{diag}(\lambda_0,\lambda_1,\ldots,\lambda_{\ell-1})
\end{equation}
Moreover, we may
choose $\lambda_0 \geq \lambda_1 \geq \cdots \geq \lambda_{\ell-1} \geq 0$.
\end{lemma}

\begin{appremark}
This is \emph{not} a diagonalization of $\eta$!  Indeed, $u$ is unitary;
$u^{-1} = u^{\dagger} \neq u^T$.  The point of this proof is to diagonalize
the $\mathbb{R}$-linear map $\eta_{\mathbb{R}} \tau$, which is symmetric.
\end{appremark}
 
\begin{proof}
Since $\eta$ is symmetric, we must have the following:
\begin{equation}
\eta_{\mathbb{R}} \ = \ 
\left( \begin{array}{rr} u & -v \\ v & u \\ \end{array} \right),
\quad \quad u=u^T, v=v^T
\end{equation}
Now if $\mu_i$ denotes scalar multplication by $i$, then clearly
$\mu_i \eta = \eta \mu_i$.  Thus
$J \eta_{\mathbb{R}} = \eta_{\mathbb{R}} J$, for $J= (\mu_i)_{\mathbb{R}}=
(-i\sigma^y)\otimes I_{\ell}$.

Now let $\tilde{\eta}$ be the associated complex anti-linear map,
i.e. $\tilde{\eta} x = \eta \overline{x}$.  Then
$\tilde{\eta}_{\mathbb{R}} = \eta_{\mathbb{R}} \tau$, and noting
that $\tau = \mbox{diag}(I_N, -I_N)$ produces
\begin{equation}
(\tilde{\eta})_{\mathbb{R}} \ = \ \eta_{\mathbb{R}} \tau \ = \ 
\left( \begin{array}{rr} u & -v \\ -v & -u \\ \end{array} \right)
\end{equation}
Hence $\tilde{\eta}_{\mathbb{R}} = \tilde{\eta}_{\mathbb{R}}^T$, i.e.
$\tilde{\eta}_{\mathbb{R}}$ is diagonalized by some matrix
orthogonal matrix.  We next argue that such an orthogonal matrix $o$ may
be chosen to be symplectic (within $Sp(\ell,\mathbb{R})$.) 
Per earlier discussion, $o \in Sp(\ell,\mathbb{R}) \cap O(2\ell)$ will then
demand $o=u_{\mathbb{R}}$ for some $u \in U(\ell)$.

\noindent
{\bf Step \# 1:}  Note that for any eigenvalue of $\tilde{\eta}_{\mathbb{R}}$,
say $\lambda$, we also have $-\lambda$ as an eigenvalue.  For
$J \tilde{\eta}_{\mathbb{R}} = - \tilde{\eta}_{\mathbb{R}} J$,
since $\tilde{\eta}$ is $\mathbb{C}$-antilinear.  Hence
$\tilde{\eta}_{\mathbb{R}} J v = - J \tilde{\eta}_{\mathbb{R}} v =
(- \lambda) Jv$, given $v \in V_\lambda \subset \mathbb{R}^{2\ell}$.

\noindent
{\bf Step \#2:}  Choose a collection of positive eigenvalues
and take $\Lambda = \mbox{diag}(\lambda_0,\ldots,\lambda_{\ell-1})$.
For the corresponding orthonormal eigenvectors
$v_0, \cdots, v_{\ell-1}$, consider the orthogonal matrix
$o = ( v_0 \cdots v_{\ell-1} Jv_0 \cdots J v_{\ell-1})$.  Then
$Jo =o J$, i.e. $o \in Sp(\ell,\mathbb{R})$.  Moreover
$o \tilde{\eta}_{\mathbb{R}} o^T = \Lambda \oplus (-\Lambda)$.
Note that without loss of generality,
$\Lambda$ reflects a choice
$\lambda_0 \geq \lambda_1 \geq \cdots \lambda_{\ell-1} \geq 0$.

\noindent
{\bf Step \#3:}  Let $u$ be the unitary matrix resulting from $o$ in
the last step.  Then the final equation demands
$u \tilde{\eta} u^\dagger = \Lambda \tau$,
i.e. $u \eta \tau u^\dagger \tau = \Lambda$, i.e. $u \eta u^T = \Lambda$.
\end{proof}

\begin{appdefinition}
\label{def:eta}
Let $\mathcal{E}= \{ x_j \}_{j=0}^{N-1}$ be an ensemble for some
density matrix $\rho$, and let $\Theta=\omega \tau$ be a time-reversal
symmetry operator.  Then we define
$\eta(\mathcal{E},\Theta)$ to be that matrix whose entries are
$[\eta(\mathcal{E},\Theta)]_{jk} = x_j^\dagger \Theta x_k =
x_j^\dagger \omega \overline{x}_k$.  We often suppress the arguments and
write $\eta=\eta(\mathcal{E},\Theta)$ when the context is clear.
\end{appdefinition}

\begin{lemma}
\label{lem:etaetabar}
We have the following basic properties of $\eta(\mathcal{E},\Theta)$.
\begin{enumerate}
\item
$\eta(\mathcal{E} \cdot u, \Theta) \ = \ 
\overline{u} \; \eta(\mathcal{E},\Theta) \; u^\dagger$.
\item
For any two ensembles $\mathcal{E}_1$, $\mathcal{E}_2$ of a given
$\rho$ of length $\ell$, we have
\begin{equation}
\mbox{spec}[ \;
\eta(\mathcal{E}_1,\Theta) \; \overline{\eta}(\mathcal{E}_1,\Theta) \; ]
\ = \ 
\mbox{spec}[ \; 
\eta(\mathcal{E}_2,\Theta)\; \overline{\eta}(\mathcal{E}_2,\Theta) \; ]
\end{equation}
\item
If $\mathcal{E}=\{ x_j \}_{j=0}^{\ell-1}$, then
$c_{2p}(x_k) = \eta_{kk}$.
\end{enumerate}
\end{lemma}

\begin{proof}
Keeping $\mathcal{E}= \{ x_j \}_{j=0}^{N-1}$, note that
$\mathcal{E} \cdot u$ is $w_j = \sum_{k=0}^{n-1} x_k e_j^\dagger u e_k$.
Thus
\begin{equation}
\begin{array}{l}
[\eta ( \mathcal{E} \cdot u, \Theta)]_{jk} \ = \ 
w_j^\dagger \omega \overline{w}_k \ = \\ 
\sum_{p=0}^{\ell-1} \sum_{q=0}^{\ell-1}
e_p^\dagger u^\dagger e_j x_p^\dagger \omega \overline{x}_q e_k^\dagger
\overline{u} e_q \ = \ \\
\sum_{p=0}^{\ell-1} \sum_{q=0}^{\ell-1} \overline{u}_{jp} 
[\eta(\mathcal{E},\Theta)]_{pq} u^\dagger_{qk} \\
\end{array}
\end{equation}
The first item results, comparing entry by entry.  For the
second item, recall that
$\mathcal{E}_2 = \mathcal{E}_1 \cdot u$ for some
unitary matrix $u$.  The third item is the definition of preconcurrence.
\end{proof}

\begin{proposition}
\label{prop:diagspec}
Fix a density matrix $\rho \in \mathbb{C}^{\ell \times \ell}$.
Suppose $\Theta$ is bosonic, so that $\omega=\omega^T$.
\begin{enumerate}
\item
\label{it:ensemble}
Then there exists a subnormalized ensemble $\mathcal{E}_0=
\{ x_j\}_{j=0}^{\ell-1}$
so that $\eta( \mathcal{E}_0,\Theta)$ is diagonal, real, and contains
only nonnegative entries.
\item
\label{it:Mrho}
Label $M(\rho) = 
[\sqrt{\rho} \omega \overline{\rho} \omega^\dagger \sqrt{\rho}]^{1/2}$.
If the length of a given ensemble $\mathcal{E}$ is $N$, then
\begin{equation}
\mbox{spec}[M(\rho)]=\mbox{spec}[\sqrt{\eta(\mathcal{E},\Theta) \ 
\overline{\eta}(\mathcal{E},\Theta)}]
\end{equation}
\end{enumerate}
\end{proposition}

\begin{proof}
We first prove Item \ref{it:ensemble}.
Fix $\mathcal{E}$ of length $\ell$ and $\Theta$, so that $\eta \in 
\mathbb{C}^{\ell \times \ell}$.  Since $\omega = \omega^T$,
we have $\eta=\eta^T$.  Hence, there exists by Lemma
\ref{lem:not_diag} a unitary matrix $u$ such that
$u \; \eta(\mathcal{E},\Theta) \; u^T = \Lambda$,
$\Lambda = \mbox{diag} (\lambda_0,\lambda_1,\cdots,\lambda_{\ell-1})$,
and $\lambda_0 \geq \lambda_1 \geq \cdots \geq \lambda_{\ell-1} \geq 0$.
Thus, $\Lambda = \eta(\mathcal{E} \cdot \overline{u},\Theta)$.

For Item \ref{it:Mrho}, let $\mathcal{E}=\{ x_j\}_{j=0}^{N-1}$.
Now by Lemma \ref{lem:etaetabar}, we may suppose without loss
of generality that $\mathcal{E}$ is a subnormalized eigenensemble,
so that $x_j^\dagger x_k = \delta_j^k$ and $\rho x_j=\nu_j x_j$
for $\{\nu_j \}_{j=0}^{N-1}$ the set of eigenvalues of $\rho$.  Form
the unitary matrix $w$ with $w e_j = x_j$, i.e.
$w = \sum_{j=0}^{N-1} x_j e_j^\dagger$.  Then
\begin{equation}
\begin{array}{lll}
w \eta \overline{\eta} w^\dagger \ &=& \bigg( \sum_{j=0}^{N-1} x_j e_j^\dagger \bigg)
\sum_{p=0}^{N-1} e_p \sqrt{\nu_p} 
x_p^\dagger \omega \overline{\rho} \omega^\dagger \\
& & \cdot \sum_{q=0}^{N-1} \sqrt{\nu_q} 
x_q e_q^\dagger \bigg( \sum_{k=0}^{N-1} x_j^\dagger e_j
\bigg) \\ &=& 
\bigg( \sum_{p=0}^{N-1} \sqrt{\nu_p}
x_p x_p^\dagger \bigg) \omega \overline{\rho}
\omega^\dagger \bigg( \sum_{q=0}^{N-1} \sqrt{\nu_q} x_q x_q^\dagger \bigg)\\ 
&=& \sqrt{\rho} \omega \overline{\rho} \omega^\dagger \sqrt{\rho} \\
\end{array}
\end{equation}
In the last line, we note that $\{ x_j \}_{j=0}^{N-1}$ is a \emph{normalized}
ensemble for $\rho$.
Hence, $\mbox{spec}(\; \eta \overline{\eta}\; ) = \mbox{spec}
(\sqrt{\rho} \omega \overline{\rho} \omega^\dagger \sqrt{\rho})$.
\end{proof}

\begin{proposition}
\label{prop:min_pos}
Let $\lambda_0 \geq \lambda_1 \geq \cdots \geq \lambda_{N-1} \geq 0$
be an ordering of the spectrum of $M(\rho)$ per Theorem
\ref{thm:uhlmann}.  Suppose in addition
that $T=\lambda_0  - \sum_{j=1}^{N-1} \lambda_j > 0$.
Then there exists a subnormalized ensemble
$\mathcal{E}_{\mbox{min}}$ of length $N$ such that
$C_{2p}(\mathcal{E}_{\mbox{min}}) = T$.
\end{proposition}

\begin{proof}
By Proposition \ref{prop:diagspec}, we produce an ensemble
$\mathcal{E}_0$ of length $N$ so that
$\eta(\mathcal{E}_0,\Theta)= \Lambda=\mbox{diag}(\lambda_0,\lambda_1,\cdots,
\lambda_{N-1})$.  Moreover, 
$\{\lambda_j \}_{j=0}^{N-1}=\mbox{spec}[M(\rho)]$, since
$\eta \overline{\eta} = \Lambda^2$.  Now label the phase matrix
$\Phi=\mbox{diag}(1,i,i,\cdots,i)$, and note that
$T=\mbox{Tr}[\overline{\Phi} \eta(\mathcal{E}_0,\Theta) \Phi^\dagger]=
\mbox{Tr}[\eta(\mathcal{E}_0 \cdot \Phi,\Theta)]$.  Moreover,
$\overline{\Phi} \eta(\mathcal{E}_0,\Theta) \Phi^\dagger$ is real.
Hence, by Lemma \ref{lem:real_trick}, there exists an orthogonal matrix
$o \in \mathbb{C}^{N \times N}$ such that
$[ o \eta(\mathcal{E}_0\cdot \Phi, \Theta) o^T]_{jj}=
\eta(\mathcal{E}_0 \cdot (\Phi o), \Theta)_{jj} = T/N$, $0 \leq j \leq N-1$.
We claim that we may now take $\mathcal{E}_{\mbox{min}} =
\mathcal{E}_0 \cdot (\Phi o)$.  Indeed, put $\mathcal{E}_{\mbox{min}}=
\{y_j\}_{j=0}^{N-1}$.  Then $y_j^\dagger \Theta y_j = 
\eta(\mathcal{E}_{\mbox{min}},\Theta)_{jj} = T/N$ for each $0 \leq j \leq
N-1$, so that
\begin{equation}
\begin{array}{l}
C_{2p}(\mathcal{E}_{\mbox{min}}) \ = \ \sum_{j=0}^{N-1} 
|y_j|^2 C_{2p}(y_j/|y_j|) = \\
\sum_{j=0}^{N-1} | y_j^\dagger \Theta y_j|
\ = \ \sum_{j=0}^{N-1} T/N \ = \ T \\
\end{array}
\end{equation}
This concludes the proof. 
\end{proof}

\begin{proposition}
\label{prop:min_zero}
Let $N=2^n$, and label
$\lambda_0 \geq \lambda_1 \geq \cdots \geq \lambda_{N-1} \geq 0$
an ordering of the spectrum of $M(\rho)$ per Theorem
\ref{thm:uhlmann}.  Suppose in addition
that $T=\lambda_0  - \sum_{j=1}^{N-1} \lambda_j \leq 0$.
Then there exists an ensemble $\mathcal{E}_{\mbox{min}}$ 
for $\rho$ such that $C_{2p}(\mathcal{E}_{\mbox{min}}) =0$.
\end{proposition}

\begin{proof}
Again, Proposition \ref{prop:diagspec} produces an ensemble
$\mathcal{E}_0$ of length $N$ so that
$\eta(\mathcal{E}_0,\Theta)= \Lambda=\mbox{diag}(\lambda_0,\lambda_1,\cdots,
\lambda_{N-1})$ with
$\{\lambda_j \}_{j=0}^{N-1}=\mbox{spec}[M(\rho)]$.
In this event, we appeal to Proposition \ref{prop:bigmess} to assert
that there must exist a collection of angles $\{\theta_j\}_{j=0}^{N-1}$
such that
$0 = | \sum_{j=0}^{N-1} \mbox{e}^{i \theta_j} \lambda_j |$.  WLOG,
take $\theta_0=0$.
Now put $\Phi=\mbox{diag}(1,\mbox{e}^{-i\theta_1/2},\cdots,
\mbox{e}^{-i \theta_{N-1}/2})$.  We then have
$\mbox{Tr}[ \overline{\Phi} \eta(\mathcal{E}_0,\Theta) \Phi^\dagger]=0$.
Recall the Hadamard computation 
$H=\frac{1}{\sqrt{2}} \sum_{j,k=0}^1 (-1)^{jk}\ket{k}\bra{j}$, 
which is unitary.  Now consider
\begin{equation}
\mbox{Tr}[H^{\otimes n} 
\overline{\Phi} \eta(\mathcal{E}_0,\Theta) \Phi^\dagger H^{\otimes n}]\
= \ \mbox{Tr}[\eta( \mathcal{E}_0 \cdot (\Phi H^{\otimes n}),\Theta)] \ = \ 0
\end{equation}
We next claim that $\mathcal{E}_{\mbox{min}} = 
\mathcal{E}_0 \cdot (\Phi H^{\otimes n}) = \{y_j\}_{j=0}^{N-1}$ is
an ensemble for $\rho$ consisting of concurrence zero states.
Indeed, put $\{z_j \}_{j=0}^{N-1} = \mathcal{E}_0 \cdot \Phi$.
Then $z_j^\dagger \Theta z_k = \mbox{e}^{i\theta_j} \lambda_j \delta_j^k$.
Now consider that due to the application of the Hadamard computation
\begin{equation}
y_j = \sum_{k=0}^{N-1} \epsilon_k z_j
\end{equation}
for each $\epsilon_k = \pm 1$.  Hence, $c_{2p}(y_j)=0$.
\end{proof}

\subsection{Minimality of 
$\mbox{max}\{0,\lambda_0-\sum_{j=1}^{N-1} \lambda_j\}$}


\begin{lemma}
\label{lem:rectw}
Let $r=\mbox{rank}(\rho) \geq \mbox{rank}[M(\rho)]$.  Consider the
first $r$, concievably nonzero eigenvalues
$\lambda_0 \geq \lambda_1 \geq \cdots \geq \lambda_r \geq 0$ of 
$M(\rho)$.  Then for any ensemble $\mathcal{E}$ of $\rho$
of arbitray length $\ell$, for $\eta = \eta(\mathcal{E},\Theta)$,
the $r$ largest eigenvalues of $\sqrt{ \eta \overline{\eta}}$
are also $\lambda_0 \geq \lambda_1 \geq \cdots \geq \lambda_{r-1} \geq 0$.
\end{lemma}

\begin{sketch}
To begin, write $\mathcal{E}$ as $\{x_j\}_{j=0}^{\ell-1}$.  Then
\(
\eta \overline{\eta} \ = \ 
\sum_{j=0}^{\ell-1} x_j^\dagger \omega \overline{\rho} \omega^\dagger
\sum_{k=0}^{\ell-1} x_k
\)
Now note that $\ell \geq r$.  Then by transitivity of the $U(\ell)$ action,
we may suppose without loss of generality that $\mathcal{E}$ is
a subnormalized eigenensemble, perhaps with trailing zero eigenvectors.
Letting $\{y_j \}_{j=0}^{\ell - 1}$ be the normalized eigenensemble,
again consider the matrix $w$ $w e_j = y_j$.  Then again for
$\{ \nu_j\}_{j=0}^r$ the nonzero eigenvalues of $\rho$,
\begin{equation}
\begin{array}{lll}
w \eta \overline{\eta} w^\dagger \ &=&  
\bigg( \sum_{j=0}^{\ell - 1} y_j e_j^\dagger  \bigg) 
\sum_{p=0}^{\ell-1} e_p \sqrt{\nu_p} y_p^\dagger 
\omega \overline{\rho} \omega^\dagger \\
& & \cdot \sum_{q=0}^{\ell-1} y_q \sqrt{\nu_q} e_q^\dagger
\bigg( \sum_{k=0}^{\ell - 1} e_j y_j^\dagger \bigg) \\
&=& \sqrt{\rho} \omega \overline{\rho} \omega^\dagger \sqrt{\rho} \\
\end{array}
\end{equation}
Although $w$ is no longer square, this statement nonetheless implies
the result on truncated spectra.
\end{sketch}

\begin{proposition}
\label{prop:lower_bound}
Let $\mathcal{E}$ be any ensemble for a density matrix $\rho$ of
length $\ell$.  Let 
$T=\mbox{max}\{0,\lambda_0-\sum_{j=1}^{\ell-1} \lambda_j\}$
for the $\lambda_j$ nonincreasing and coinciding with the
spectrum of $M(\rho)=(\sqrt{\rho} \omega \overline{\rho} \omega^\dagger
\sqrt{\rho})^{1/2}$.  Then $C_{2p}(\mathcal{E}) \geq T$.
\end{proposition}

\begin{proof}
Fix $\eta = \eta(\mathcal{E},\Theta)$.
We recall by Lemma \ref{lem:not_diag} that there exists some unitary
matrix $u$ so that $u \eta u^T = \Lambda$.
By Lemma \ref{lem:rectw}, we have
$\Lambda = \mbox{diag}(\lambda_0,\lambda_1,\lambda_2,
\cdots,\lambda_{\ell-1})$ for
$\lambda_0 \geq \lambda_1 \geq \cdots \geq \lambda_{\ell-1}\geq 0$
the $\ell$ largest eigenvalues of
$M(\rho)=
(\sqrt{\rho} \omega \overline{\rho} \omega^\dagger \sqrt{\rho})^{1/2}$.
Moreover every nonzero eigenvalue of $M(\rho)$ appears
within the set of the $\ell$ largest.

We first consider the case that $T=
\lambda_0 - \sum_{j=1}^{\ell} \lambda_j \geq 0$.
Now note that by the Definition \ref{def:eta} for $\eta$, we have
$C_{2p}(\mathcal{E}) = \sum_{j=0}^{\ell-1} |\eta_{jj}|$.  Moreover,
$\eta_{jk}=(u \Lambda u^T)_{jk} = \sum_{p=0}^{\ell-1} u_{jp} \lambda_p u_{kp}$,
so that
$\eta_{jj}=\sum_{p=0}^{\ell-1} u_{jp}^2 \lambda_p$.  The result then follows
from the Schwarz inequality, given $\sum_{p=0}^{\ell-1} |u_{jp}|^2=1$:
\vbox{
\begin{equation}
\begin{array}{lcl}
C_{2p}(\mathcal{E}) & = & \sum_{j=0}^{\ell-1} |\eta_{jj}| \\
& = & 
\sum_{j=0}^{\ell-1} \bigg| \sum_{p=0}^{\ell-1} \sum_{j=1}^{\ell-1}
u_{jp}^2 \lambda_p \bigg| \\
& \geq & \lambda_0 - \sum_{p=0}^{\ell-1} \lambda_p 
\bigg| \sum_{j=1}^{\ell-1} u_{jp}^2 \bigg| \\
& \geq & \lambda_0 - \sum_{p=1}^{\ell-1} \lambda_p \\
\end{array}
\end{equation}
}
This concludes the proof for $\lambda_0 \geq \sum_{j=1}^{\ell-1} \lambda_j$.

Thus, suppose $T=0$.  Then the statement is vacuous, since always
$C_{2p}(\mathcal{E}) \geq 0$.  This concludes the proof.
\end{proof}


\begin{thebibliography}{9}

\bibitem{Nielsen}  T.J.~Osborne and M.A.~Nielsen, Phys. Rev. A {\bf 66}, 032110 (2002).

\bibitem{Osterloh} A.~Osterloh, L.~Amico, G.~Falci, and R.~Fazio, Nature (London), {\bf 416}, 608 (2002).

\bibitem{GVidal:03} G.~Vidal, J.I.~Latorre, E.~Rico, and A.~Kitaev, Phys. Rev. Lett., {\bf 90}, 227902 (2003).

\bibitem{Verstraete:04} F.~Verstraete, M.A.~Martin-Delgado, J.I.~Cirac, Phys. Rev. Lett. {\bf 92}, 087201 (2004).

\bibitem{Barnum:03}  H.~Barnum, E.~Knill, G.~Ortiz, and L.~Viola, Phys. Rev. A {\bf 68}, 032308 (2003).

\bibitem{Somma:04} R.~Somma, G.~Ortiz, H.~Barnum, E.~Knill, and L.~Viola, quant-ph/0403035.

\bibitem{Wootters:98} W.K.~Wootters, Phys. Rev. Lett. {\bf 80}, 2245 (1998).

\bibitem{Wong:01}
A.~Wong and N.~Christensen, Phys. Rev. A { \bf 63}, 044301 (2001).

\bibitem{BB:03} S.S.~Bullock and G.K.~Brennen, J. Math. Phys {\bf 45}, 2447 (2004).

\bibitem{BBOL:04} S.S.~Bullock, G.K.~Brennen, D.P.~O'Leary, quant-ph/0402051.

\bibitem{Lieb:61}  E.~Lieb, T.~Schultz, and D.~Mattis, Ann. of Phys. {\bf 16}, 407 (1961).

\bibitem{Lieb:62} E.~Lieb and D.C.~Mattis, J. Math. Phys. {\bf 3}, 749 (1962).

\bibitem{Katsura} S.~Katsura, Phys. Rev. {\bf 127}, 1508 (1962).

\bibitem{Wang:02} X.~Wang, Phys. Rev. A {\bf 66}, 044305 (2002).

\bibitem{Teich:03} G.~Jaeger, {\it et al.}, Phys. Rev. A {\bf 67}, 032307 (2003).

\bibitem{Duan:03} L.-M.~Duan, E.~Demler, and M.D.~Lukin, Phys. Rev. Lett. {\bf 91}, 090402 (2003). 

\bibitem{NBS} Handbook of Mathematical Functions, M. Abramowitz and L.A. Stegun (National Bureau of Standards, 1964), pg. 498.


\bibitem{Brennen:03} G.K.~Brennen and J.E.~Williams, Phys. Rev. A {\bf 68}, 042311 (2003).

\bibitem{Khaneja:02} N.~Khaneja and S.J.~Glaser, Phys. Rev. A {\bf 66}, 060301(R) (2002).

\bibitem{Benjamin:03} S.C.~Benjamin and S.~Bose, Phys. Rev. Lett. {\bf 90}, 247901 (2003).

\bibitem{Helgason:01}
S.~Helgason,
\newblock {\em Differential geometry, {L}ie groups, and symmetric spaces},
  volume~34.
\newblock American Mathematical Society, Providence, RI, graduate studies in
  mathematics, (corrected reprint of the 1978 original) edition,  (2001).

\bibitem{HughstonEtAl:93}
L.~Hughston, R.~Jozsa, and W.K.~Wootters,
Phys. Lett. A {\bf 183}, 14 (1993).

\bibitem{Uhlmann:00}
A.~Uhlmann, 
Phys. Rev. A {\bf 62}, 032307 (2000).


\end{thebibliography}
\end{document}